\documentclass[a4paper,amsfonts, amssymb, amsmath, reprint, showkeys,twoside,onecolumn,pra]{revtex4-1}

\usepackage[english]{babel}
\usepackage[utf8]{inputenc}
\usepackage{xcolor}
\usepackage{color}
\usepackage{amssymb}
\usepackage[colorinlistoftodos, color=green!40, prependcaption]{todonotes}
\usepackage[pdftex, pdftitle={Article}, pdfauthor={Author}]{hyperref} 
\usepackage[version=4]{mhchem}
\usepackage{physics}
\usepackage{hyperref}
\usepackage{url}
\usepackage{float}

\usepackage{multirow}
\usepackage{xcolor}

\begin{document}

\title{The pros and cons of using deep reinforcement learning or genetic algorithms to design control schemes for quantum state transfer on qubit chains}

\author{Sofía Perón Santana$^{(1)}$$^{(2)}$}
\author{Ariel Fiuri$^{(1)}$}
\author{Martín Domínguez$^{(1)}$}
\author{Omar Osenda$^{(1)}$$^{(2)}$}

\email[Correspondence email address: ]{sofia.peron@mi.unc.edu.ar}
\affiliation{${(1)}$ Facultad de Matemática, Astronomía, Física y Computación, Universidad Nacional de Córdoba, Av. Medina Allende s/n, Ciudad Universitaria, CP:X5000HUA Córdoba, Argentina, and 
${(2)}$ Instituto de Física Enrique Gaviola (CONICET-UNC) and }

\date{today}

\begin{abstract}
In recent years, control methods based on different optimization techniques have shed light on the possibilities of processing information in many quantum systems. When exploring the transmission of quantum states, faster transmission times are mandatory to avoid the deleterious effects of multiple sources of decoherence that spoil the transmission process. In particular, using Reinforcement Learning to devise sequences of step-wise external controls provides good transfer policies at short transmission times. We present two approaches to control the transmission of quantum states in qubit chains using external controls to force the dynamical evolution of the chain state. The first approach relies on the well-known Genetic Algorithm to generate a sequence of external controls, while the second approach uses a variant of Reinforcement Learning. The Genetic algorithm achieves excellent transmission fidelity at as short transmission times as Reinforcement Learning, surpassing the fidelities achieved by the latter method. Nevertheless, the Reinforcement Learning method offers robust control policies when the control pulses are noisy enough, owing to an imperfect timing of the pulses, deficient control devices, or other sources of phase decoherence. We present the regime where each method is best suited to control the transmission of arbitrary qubit states. 
\end{abstract}

\maketitle

\section{Introduction}

Controlling the time evolution of the quantum state \cite{Bukov2018,Sivak2022,Niu2019} is the fulcrum on which quantum technologies, such as computing \cite{Burgarth2010}, quantum state transfer \cite{Heule2010, Coden2020,Zhang2016,SerraPLA,Chan2023}, state preparation \cite{Stefanatos2019}, and cryptography, operate. The exquisite level of control needed to perform quantum algorithms that provide sensible answers is, at present, beyond the accuracy levels required to perform large-scale calculations, although the frontier between what is doable and not is moving by the day. 

Optimal Control Theory is a method commonly employed to construct external control pulses for quantum \cite{Werschnik2007,Putaja2010} and classical systems, owing to its high accuracy and clear physical interpretation, which is grounded in the variational principle. Nevertheless, its use is far from direct, and when the number of actuators increases, the numerical cost can become prohibitive. There is a growing demand for methods that provide control policies for the increasing number of platforms where quantum information is processed \cite{Bukov2018,Sivak2022,Niu2019}. From the design of control pulses for single-quantum gates to the control of shallow quantum circuits \cite{Perrier2020}, other machine learning methods are being developed or adapted for quantum systems \cite{Zhang2018,Peron2025,Mortimer2021,Loft2016}. 

In the long run, controlling open quantum demands methods resilient against decoherence and, ideally, faulty implementations of the control pulses \cite{Weidner2025,Keele2022,Sakuldee2024}. This goal suggests that those control methods that rely on variable environments to train an agent should be up to the task. Reinforcement learning methods naturally incorporate the described ingredients. The numerous implementations available for studying a given problem should allow choosing the most appropriate Reinforcement Learning version for the investigation at hand \cite{Li2023,Sutton1998,Murphy2025}. Regrettably, there is a lot of trial and error to overcome before gaining insight into which traits of a given version of the Reinforcement Learning method are indispensable and which are superfluous \cite{Bukov2018,Sivak2022,Porotti2022,Niu2019,Rahimi2025}. 

Among many problems, reinforcement learning methods allow the preparation of random pure states in systems of up to seven qubits using different control sets \cite{Rahimi2025}, enhance the transmission of quantum states in qubit chains, and improve the performance of quantum gates. Although the transmission of quantum states achieved using Reinforcement Learning with Deep Q-Networks with an {$\epsilon$} greedy policy for short to moderate qubit chain lengths is not of a remarkable quality \cite{Zhang2018}. Nevertheless, it shows that Reinforcement Learning could provide control sequences compatible with fast transmission. Training reinforcement Learning models without noise or some disturbances does not result in a well-trained model \cite{Tobin2017,Weng2019}. Indeed, for an open quantum system or one with faulty controls,  training with noise is a matter of using it as an advantage, a problem, and transforming it into an opportunity. A trained model provides robust control sequences for a broad spectrum of situations \cite{Gruosso2021, Wu2024}, assuming the training included a large set of possibilities. Otherwise, faced with a slightly different problem, the model will not provide a sensible answer. 

On the other hand, in general, Genetic Algorithms \cite{genetico} work best for systems without fluctuations and provide very good to excellent solutions to problems when a given quantity needs to be optimised. In most situations, the resulting controls fail when used on systems with fluctuations or faulty implementations. Qubit chains with site-dependent interactions, designed using the Genetic Algorithm (GA), have demonstrated excellent transmission performance \cite{Peron2025, Mortimer2021,Bezaz2025,Nelmes2025,Faria2025}. The preparation of particular sets of quantum states using GA is possible. All in all, there is an increasing field of applications. In the most common cases analysed using the Genetic Algorithm, the setting results in tuning the parameters of a time-independent Hamiltonian. Nevertheless, more recently, the design of control pulses has been considered.

In this work, we aim to compare the performance of both the Genetic Algorithm and the Reinforcement Algorithm when applied to the problem of transmitting quantum states over homogeneous qubit chains. The study of quantum state transfer with \cite{Xie2023,Maleki2021,Coden2020,Zhang2016,Randles2024,Xiang2024} and without external controls \cite{Serra2022,Mograby2021,Wang2016,Serra2024,Estarellas2017,Feldman2024,Romero2024,Stolze2014}, using qubit chains with site-dependent interactions \cite{Kostak2007,Peron2025,FerronPS,Zwick2015} or homogeneous ones \cite{Kandel2019}, has led to numerous protocols verified in different experimental setups, such as optical systems \cite{Chapman2016,Baum2021}, superconducting qubit processors \cite{Li2018}, magnetic systems \cite{nuclear-spin-chain-1,nuclear-spin-chain-2,nuclear-spin-chain-3}, cold atoms \cite{Loft2016,Banchi2011prl}, quantum dot-based qubit chains \cite{Yang2010,quantum-dot-chain,Faroq2015,Kandel2021,Martins2017} and  others \cite{Yue2024}. As such, quantum state transfer is one of the tasks instrumental for assessing the utility and reach of every control technique \cite{Kay2010}.

GA is best suited for problems without fluctuations, whereas DRL algorithms find their best applications in problems with fluctuating environments. In situations with fluctuating environments, it is possible to validate the trained model \cite{Gruosso2021,Wu2024}. Even starting the physical system under study in the same initial state results in different answers, which, for a properly trained model, would produce high-quality quantum state transfer. So, using an appropriate noise model, we can compare the performance of control sequences designed using both types of algorithms to drive quantum state transfer in qubit chains. It makes sense to study the generation of control sequences and their properties first without fluctuations, for both types of algorithms, and then with fluctuations, paying attention to the appearance of new phenomena.

We organised the paper as follows. In Section \ref{sec-model-and-methods}, we present the qubit chain model Hamiltonian and details about the algorithms employed in the paper. Section \ref{sec-ga} contains the results for the transmission probability obtained using the Genetic Algorithm. Section \ref{sec-reinforcement-learning} presents results about control strategies found using Reinforcement Learning and a comparison with the results obtained using the Genetic Algorithm. Section \ref{sec-noisy-control} is devoted to the effects of fluctuations when they are present on the time evolution of the quantum state. Here, we train and validate DQN models that generate control sequences to drive quantum state transfer. At the same time, by using control sequences generated by the GA without fluctuations, we test their robustness by adding fluctuations to those sequences. The fluctuations model employed in both the GA and DQN algorithms is the same. Finally, in Section \ref{sec-discussion-and-conclusions}, we summarise our results and present our conclusions. 

To make the presentation lighter and enhance the readability of the manuscript, we defer most technical details, tables with parameters, and precise definitions of some quantities to the Appendices. Appendix~\ref{ap-genetic-details} contains what is understood by mutation type, crossover type, number of generations, saturation and other quantities that modify the behaviour of the GA. Most of the Figures which include data obtained using control sequences generated through the GA use the parameters listed in the Table included in this Appendix. Appendix~\ref{ap-RL-details} contains a pseudo-code detailing the main ingredients of the Deep Q-network algorithm employed in this work. DQN is a variant of Deep Reinforcement Learning methods, which depends on artificial neural networks to approximate the transition probabilities needed for the underlying Markov Decision Process. The architecture of the networks, learning rates, reward and other necessary quantities included in the pseudo-code are the subject of this Appendix. Appendices~\ref{ap-histograms} and \ref{ap-largedims} are devoted to characterising the control sequences obtained using the GA, in particular, what control actions are involved in constructing successful control sequences and the quality of the quantum state transfer obtained using a fixed set of parameters for larger and larger systems. 



\section{Model and methods}
\label{sec-model-and-methods}

We will study the transmission of quantum states over qubit chains equipped with the $XX$ Hamiltonian,

\begin{equation}\label{ec-xx-hamiltonian}
H = - \sum_{i=1}^{N-1} J \left( \sigma_{i}^x \sigma_{i+1}^x +  \sigma_{i}^y \sigma_{i+1}^y \right) + \sum_{i=1}^{N} h_i \sigma_i^z .
\end{equation}

The "external fields" $\lbrace h_i \rbrace$ provide the control needed to force the time evolution from the initial state to the desired target. Different combinations of fields could be applied and analysed. Nevertheless,  simple combinations are preferable. On modern quantum processors, applying a control pulse on the $z$ direction to a given qubit is equivalent to using a phase gate on it. If the qubits have magnetic dipolar momentum, the control pulses are, effectively, external magnetic fields, but this is not the most general case. For instance, in superconducting qubits, the phase gate is implemented via a radio frequency pulse. The Hamiltonian in Eq.~\eqref{ec-xx-hamiltonian} commutes with the total magnetization, allowing the study of the transmission of quantum states in subspaces with a fixed number of excitations. 

To control the time evolution of the quantum state, we use control functions $\lbrace h_i\rbrace$ that are piecewise constant functions of time. Over a given time interval, only a particular set of control functions is non-zero. So, for the time interval $\tau_j$, the control term of the Hamiltonian is given by
\begin{equation}
\sum_{h_i(\tau_j)\neq 0} h_i(\tau_j) \sigma_i^z .
\end{equation}

Calling $H_j$ to the sum of the $XX$ Hamiltonian and the control term valid for the $\tau_j$ interval, we get that
\begin{equation}
H_j = - \sum_{i=1}^{N-1} J \left( \sigma_{i} \sigma_{i+1} +  \sigma_{i} \sigma_{i+1}\right) + \sum_{h_i(\tau_j)\neq 0} h_i(\tau_j) \sigma_i^z  .
\end{equation} 
So, we assume that the quantum state transfer takes place at a transmission time $T \in \left[ 0, t \right]$, and 
\begin{equation}
t= \tau_1 + \tau_2 + \ldots + \tau_{n-1} +\tau_{n} .
\end{equation}
Then, the vector state of the qubit chain at time $t$ is given by
\begin{equation}\label{ec-evolution-operator}
\psi(t)= \exp{(i H_n \tau_n)} \exp{(i H_{n-1 }\tau_{n-1})} \ldots \exp{(i H_2 \tau_2)} \exp{(i H_1 \tau_1)} \psi(0)  = U \psi(0).
\end{equation}
So, controlling the quantum state transfer involves finding a sequence of controls
\begin{equation}
\lbrace h_i(\tau_j) \rbrace, \quad \forall j=1,2,\ldots , n
\end{equation}
such that the initial state, dwelling at one extreme of the qubit chain at $t=0$, is transferred with as much fidelity as possible at the other extreme of the chain. At this stage, the transmission protocol changes depending on the number of qubits used to encode the initial state and other details. Nevertheless,  we will focus on the simplest one, the transmission of an arbitrary one-qubit state. 

The probability of transmission of a one-qubit arbitrary quantum state, at time $t$, on the computational basis can be written as
\begin{equation}\label{ec-transmission-probability}
P(t) = | \langle \mathbf{N} | U(t) | \mathbf{1} \rangle |^2 , 
\end{equation}
where $U(t)$ is the evolution operator in Eq.~\eqref{ec-evolution-operator}, $| \mathbf{1} \rangle= |1\otimes o \otimes 0 \otimes \ldots\otimes 0\rangle$ and 
$| \mathbf{N} \rangle= |0\otimes 0 \otimes 0 \otimes \ldots\otimes 1\rangle$ are the one-excitation states with the excitation located at the first and last sites of the chain, respectively. The averaged fidelity of transmission is a simple function of $P(t)$,
\begin{equation}
f(t) = \frac{P(t)}{6} + \frac{\sqrt{P(t)} \cos{\gamma}}{3} +\frac{1}{2} ,
\end{equation}
where the average considers all the possible initial states, and $\gamma$ is a phase difference, which satisfies that $\cos{\gamma}=1$ by applying an appropriate uniform and time-independent magnetic field to the chain. 

While in modern quantum superconducting processors (see Reference~\cite{Zuchongzhi2025} and references terein), it is possible to control the time evolution of the quantum state by applying multiple one- and two-qubit gates simultaneously, we will use only one-qubit controls acting on a single qubit or a few of them. Mainly, we will consider that only a single $h_i(\tau_j)$ is non-zero for a given time interval $\tau_j$.  

\begin{figure}[H]
\includegraphics[width=0.7\linewidth]{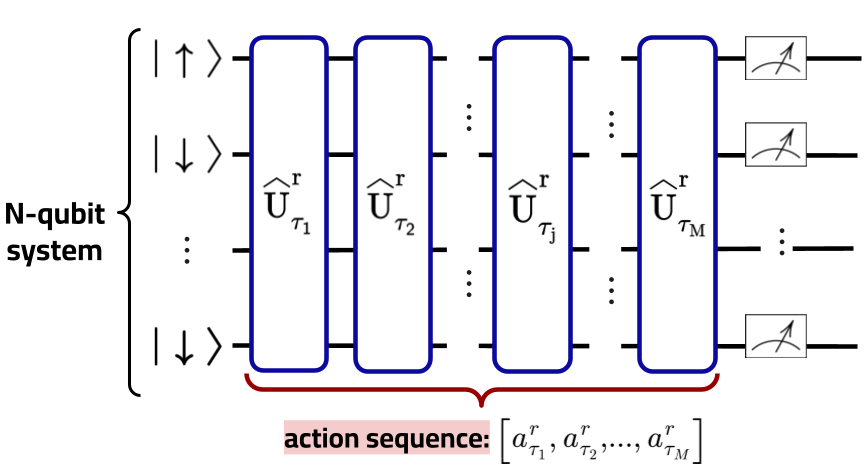}
\caption{The cartoon in the figure depicts a system of $N$ qubits and its time evolution. The initial state, shown at the leftmost extreme of the cartoon, corresponds to a one-excitation quantum state. The step-wise evolution operator for a given interval, $U_k = U(\tau_k)$, acts over all the qubits and is shown as a gate that applies over the whole system. The specific form of each evolution operator depends on the controls operating on the corresponding time interval. }\label{fig-cartoon-circuit}
\end{figure}

We call an {\em action} the set of controls $h_i(\tau_j)$ that are different from zero for a given time interval $\tau_j$. The objective of the Genetic Algorithm and the Reinforcement Learning algorithm is to determine sequences of actions resulting in high-quality quantum state transfer. The cartoon in Figure~\ref{fig-cartoon-circuit} depicts the time evolution of the quantum state of a system of $N$ qubits for a given sequence of actions. 

Without going into too much detail, it is worth mentioning the traits that enable formulating the quantum state transfer problem in a form amenable to study using GA and RL techniques. We define the temporal step $\Delta t$ so, for all $k$, $\tau_k= \Delta t$. The actions belong to an {\em action space}, $A$, and there is an {\em state space}, $S$. The state of the system is a vector with $2N$ real coefficients derived from the $N$ complex quantum vector state. All the possible states form the {\em state space}. Besides, it is useful to introduce the transition probability, which is the probability that a given action $a$ drives the system from the state $s(t)$ to the state $s^{\prime}(t+\Delta t)$. There are numerous ways to denote the transition probability, but we choose the notation $P_a(s,s^{\prime})$. Applying an action that drives the state from $s$ to $s^{\prime}$ also produces an immediate reward, $R_a(s,s^{\prime})$, which quantifies how promising the action is in terms of the task assigned to the agent. The tuple $(S, A, P_a, R_a)$ determines a Markov Decision Process (MDP) \cite{Sutton1998,Li2023,Nair2015}. 

An objective of the MDP is the optimisation of the cumulative reward function through a policy, usually denoted by $\pi$. The policy that maximises the cumulative reward is called an optimal policy \cite{Sutton1998,Li2023,Nair2015}. Deep Reinforcement Learning methods aim to {\em learn} the optimal policy by training an agent and are preferred when the dimension of the state space and the action space becomes too large. As we will see in more detail later in this work, the GA determines a sequence of actions which is the chromosome of the fittest individual in a population of control sequences. On the other hand, the DRL chosen by us to study quantum state transfer, the Deep Q-network, is an instance of an MDP. The agent of the DQN learns to recommend sequences. 

Finally, when comparing the results obtained using the GA and DQN methods, we enforce some fairness criteria, the same action sets, halting criteria, the computational cost evaluated in a reference computational system, and the best solution corresponds to the maximum value of the transmission probability found in a given time interval $\left[0, t_{max}\right]$. In this case, the best solution offered by the DQN method appears in any training episode, each one lasting from $t=0$ to $t=t_{max}$. These points are the subjects of Sections III and IV. We devote Section V to obtaining a well-trained agent that recommends control sequences that endure fluctuating dynamics.

\section{Control policies designed using Genetic Algorithm }
\label{sec-ga}

The Genetic algorithm is biologically motivated, and its rules mimic rules found in the formation of new populations with chromosomes that come from the genetic pool of previous populations, with the addition of random mutations \cite{genetico}. The purpose of the algorithm is to increase the fitness of the population, just as is the case in biology. The fitness quantifies how much a given chromosome is adapted and evaluated for all the individuals that form a population. As we consider a discrete algorithm, the offspring of a given population, together with a number of the parents, form the new generation of the population. Central to the successful execution of the algorithm are the number of parents allowed to mate, typically those with the highest fitness, the genes each parent contributes to the chromosome of a given offspring, and the number of parents, among other factors. 

All the quantities described above form the set of hyper-parameters of the algorithm \cite{genetico,PyGAD}, and their values affect the size and characteristics of the space where it searches for the optimal solution. In Appendix~\ref{ap-genetic-details}, we present further details about the Genetic Algorithm and carefully describe the rules that produce the genetic pool of the successive generations. The fitness function requires further analysis since different choices could lead to quantitative and qualitatively distinct solutions. We will return to this point once we formulate the search for the optimal quantum state transfer problem in a problem adapted for the Genetic Algorithm. 

As stated in the previous Section, we seek a control sequence which results in high values of the probability of transmission. So, it is reasonable to consider the chromosome of an individual to be the set of actions that drives the dynamical evolution from the initial state up to the final state. Each gene on a chromosome corresponds to an action. The initial population required by the algorithm comprises the chromosomes of all individuals in the population, and each gene corresponds to a randomly selected action from the set of possible actions.

\begin{figure}[H]
\includegraphics[width=0.9\linewidth]{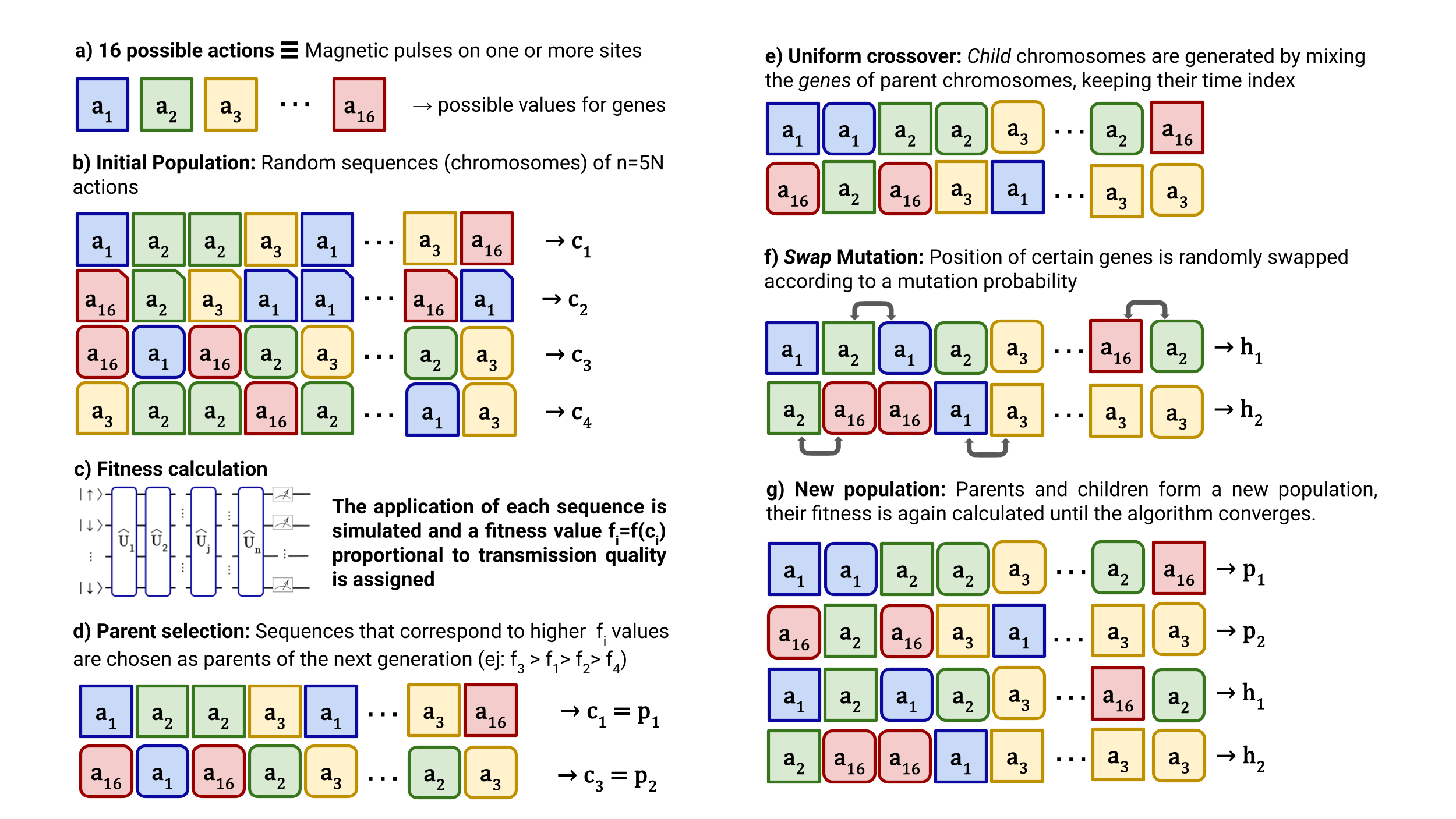}
\caption{The cartoon in the Figure presents the main ingredients of the Genetic Algorithm. a) The sixteen possible actions, each of which can appear on a control sequence at any position in it. b) An initial population of four individuals, each one endowed with its own chromosome. The chromosome contains the genes, each one an action. c) For each sequence $c_i$ (or chromosome), the algorithm calculates its fitness $f_i$, which corresponds to the maximum value attained by the transmission probability along the time interval $\sum_k \tau_k$. d) The algorithm selects a set of fittest individuals (in this case, two parents). e) and f) the chromosomes of the new members of the population result from the application of two rules, uniform crossover and swap mutation. g) The new population contains the set of parents and their offspring.  }\label{fig-genetic-details}
\end{figure}

The cartoon in Figure~\ref{fig-genetic-details} depicts the main ingredients of our implementation of the genetic algorithm.  

The cartoon depicts a case where a) there are sixteen possible actions and b) an initial population with four individuals, each one with its corresponding chromosome. The length of the chromosome is not specified, but it is the same for all the individuals. The fitness evaluation takes place in c). d) After the fitness evaluation, only the parents with the highest fitness remain in the population. The next generation individuals comprise the two parents and two offspring, whose chromosomes result from e) the {\em crossover} of the parents' genes. Sometimes, a mutation occurs that {\em swaps} the position of some genes in the chromosome of an individual.  

For the fitness function,  we choose the maximum value attained by the transmission probability over a time interval $\left[ 0, t \right]$, where $t \lesssim N$, since we seek fast transmission times. In other optimisation methods, such as Optimal Control Theory, the time at which transmission probability attains its maximum value is chosen {\em a priori}. Our results show that both the Genetic and Reinforcement Algorithms find control policies compatible with transmission times near the shortest times allowed by the Quantum Speed Limit \cite{Coden2020,Zhang2018,Xie2023}. However, the quality of the transmission differs.

In this Section, we consider two types of actions. Following the idea that, with current technology, it is possible to apply one-qubit gates to all the qubits in a chain. The action-by-site approach considers that only one field $h_i=h$ is non-zero in any particular time-bin $\tau_j$, where $h$ is the strength of the control. As we also consider that $\tau_j=\Delta t = t/n$, where $n$ is the number of controls applied (or the number of phase gates), the phase gate applied to, say, the $k$-th qubit is given by 
\begin{equation}
U_{phase} = e^{i h \Delta t \sigma_k^z } .
\end{equation}
In actual implementations, the values of $h$ and $\Delta t$ are bound by what is doable with the control means at disposal. Nevertheless, a practical bound follows from the fact that a phase gate is periodic. So, in an actual case, the time interval $\Delta t$ should be large enough to allow the control system to apply any phase gate such that $h \Delta t \leq 2\pi$. As a consequence, the number of one-qubit gates needed to explore the time interval $\left[ 0, t \right]$ is given by
\begin{equation}
N_q = \frac{t}{\Delta t} =  \frac{3}{4} \frac{N}{\Delta t} ,
\end{equation}
where we assume that the transmission time $t \lesssim 3N/4 $. 

\begin{figure}[H]
\begin{center}
\includegraphics[width=0.6\linewidth]{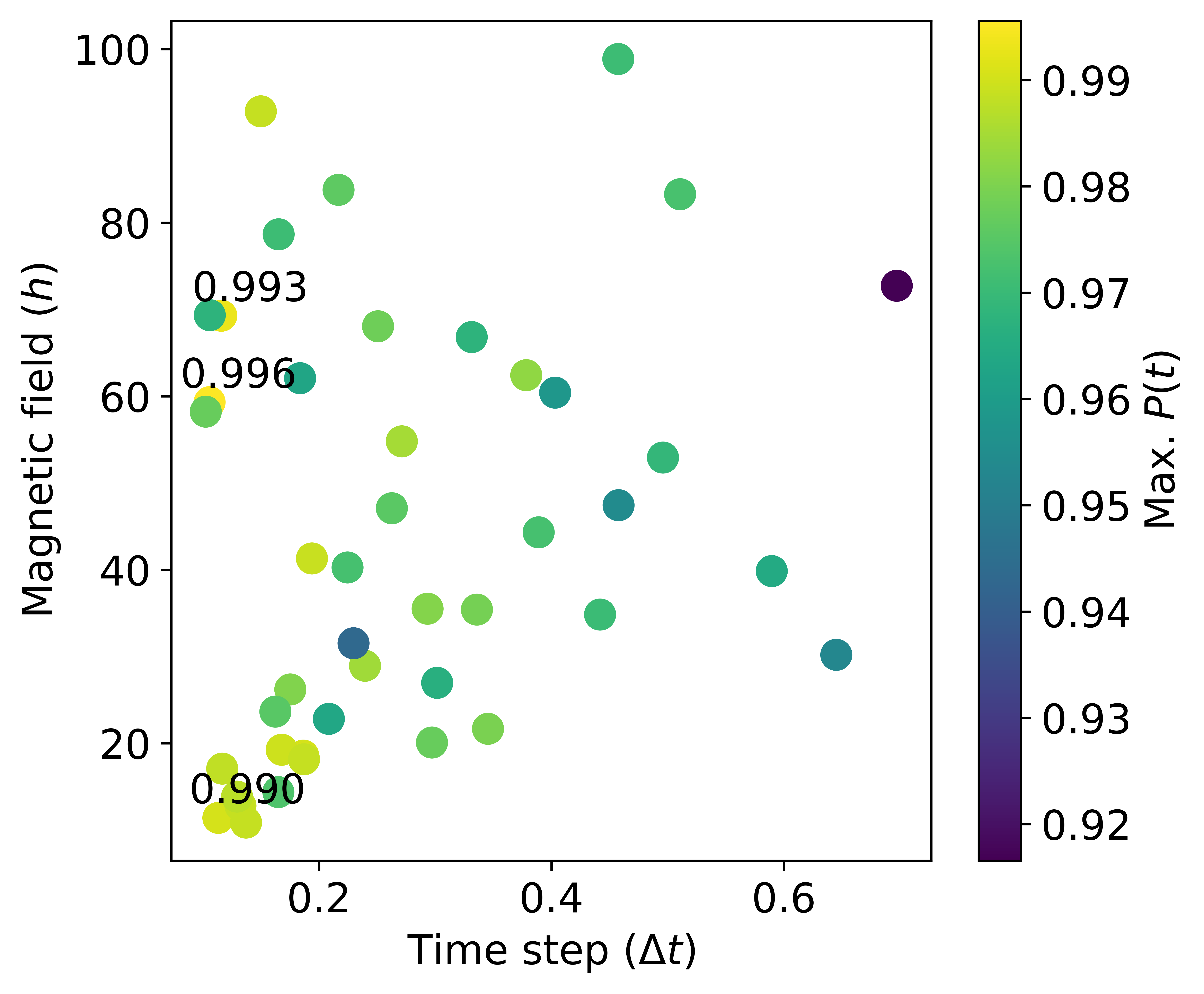}
\caption{The figure shows the maximum value of transmission probability achieved by the genetic algorithm for a chain of $N=64$ spins for different values of $h$ and $\Delta t$.  }\label{fig-figure1}
\end{center}
\end{figure}

Having the previous analysis in mind, we tested the performance of the GA using different combinations of the parameters $h$ and $\Delta t$ and summarised our findings in Figure~\ref{fig-figure1}. As the data points indicate, the GA identifies action sequences that result in high-quality quantum state transfer for many different values of both parameters. 

For comparison purposes, we also implemented the control actions proposed by Zhang and collaborators in Reference~\cite{Zhang2018}, where an action acts upon up to three qubits in one or the other extremes of the chain. In this case, the number of actions is the same irrespective of the chain length. We include this set of actions in Appendix~\ref{ap-genetic-details}. Figure~\ref{fig-fid-max-best-ga} shows the performance of the GA using both sets of actions. The results obtained using the action-by-site scheme outperform those obtained using the set of actions by Zhang and collaborators, even when the number of actions is similar. In what follows, we use $\Delta t=0.15$, $J=1$ and $h=100$, to meet our comparison criteria. 

\begin{figure}[H]
\begin{center}
\includegraphics[width=0.99\linewidth]{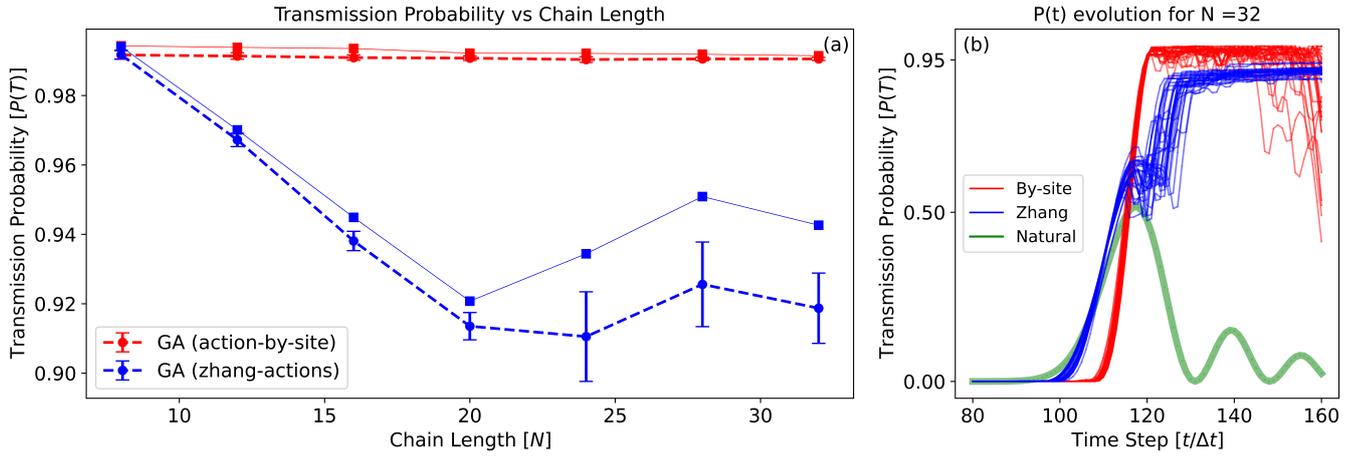}
\caption{ \label{fig-fid-max-best-ga} The figure summarises the effect of the initial conditions on the values obtained for the transmission probability. The GA is a biased random search algorithm; consequently, it is always advisable to study the variability of the results obtained. Panel a) shows the maximum value obtained for the transmission probability considering $30$ different initial populations. The solid dots correspond to the values of the transmission probability, the circular dots correspond to the maximum over all the realisations of the initial populations, and the square dots to the value obtained by averaging the maxima of the realisations. The site-by-site action set results are much stable than the action set with a fixed number of elements. Note the different sizes of the bar errors, which correspond to one standard deviation. We obtained the data points by imposing convergence criteria, by asking each algorithm to halt after the transmission probability reaches a target figure of $0.99$, after a total of $1000$ generations or after $30$ generations without an increment in fitness. Panel b) shows the time evolution obtained using the control sequences whose transmission probabilities appear in panel a). }
\end{center}
\end{figure}

Panel a) of Figure~\ref{fig-fid-max-best-ga} shows the best values found for the transmission probability using the two sets of actions described so far. The data points obtained using actions-by-site show that the quality of transmission remains remarkably high for all the chain lengths included in the figure; instead, the quality of the transmission strongly decays when the algorithm employs the action set proposed by Zhang and coworkers. In both cases, the solid square dots correspond to the best values found running the algorithm for $M=10$ different initial populations. The solid circular dots correspond to the averaged values of the maximum transmission probabilities found for each realisation of the initial populations. Not only are the transmission probabilities larger for the action-by-site actions set, but their standard deviations are much smaller. Panel b) shows the time evolution of the transmission probabilities calculated using the control sequences designed for the GA, corresponding to the sequences that result in the probabilities shown in panel a). The many red and blue curves correspond to the sequences of both action sets. The green curve shows the unforced autonomous evolution of the transmission probability. 

There are some remarkable facts appearing after a close examination of panel b). First, the action set proposed by Zhang generates action control sequences for which the transmission probability follows the natural evolution quite closely, up to the transmission time where the transmission probability produced by the natural evolution attains its maximum value. After this time, the control sequence prevents $P(t)$ from decaying, driving the time evolution up to a new maximum value, at a transmission time that is slightly larger than the time where the transmission probability corresponding to the natural evolution attains its maximum, which is compatible with the Quantum Speed Limit. The action-by-site action set strategy generates control sequences such that the transmission probabilities do not follow the natural time evolution so closely, starting their growth a bit later than the growth shown by the natural time evolution. Nevertheless, the growth of the transmission probability forced by these control sequences outgrows the other control strategy and reaches a very high maximum at practically the quantum Speed Limit time. Finally, the two types of control sequences almost freeze the time evolution of the state, forcing the excitation to remain in the last site of the chain. 



\section{Control policies designed using Reinforcement Learning}
\label{sec-reinforcement-learning}

Reinforcement learning is another machine learning method which proceeds by using a trial-and-error strategy. There are numerous variants, depending on the Markov decision process (MDP) formulation (state, action, and reward), as well as on-policy vs. off-policy, value-based vs. policy-based, among other possibilities \cite{Sutton1998,Li2023}. Also crucial in determining which variant of RL is appropriate for a given problem are the characteristics of the data used to learn. The book by Shengbo Eben Li \cite{Li2023} lists several  Reinforcement Learning algorithms and Deep Reinforcement Learning (DRL) algorithms, describing their advantages and drawbacks. For instance, the Proximal Policy Optimisation (PPO) algorithm is used on Reference ~\cite{Rahimi2025} to find control sequences for the state preparation problem, and Zhang and collaborators \cite{Zhang2018} used a Deep Q-Network algorithm (DQN) to improve the transmission of quantum states and reveal the constraints imposed by the Quantum Speed Limit to the efficiency of the transmission.  DRL algorithms are beneficial, in addition to being a natural choice, when the product of the state space and the action space is huge, or the state space is continuous. In this case, the network's role is to approximate the probability density of transitions between states given the set of actions that force temporal evolution. 

In this Section, we also use DQN to design control sequences using the two sets of actions described so far. We will keep the description of the DQN algorithm to a minimum, since both References~\cite{Sutton1998,Li2023} describe it extensively. Nevertheless, we include definitions of the hyperparameters used by the algorithm and other quantities in Appendix~\ref{ap-RL-details}, and use the main text to present the relationship between the quantum state transfer problem and the characteristic functions that enter into the DQN algorithm. 

For our purposes, it is sufficient to say that the DQN algorithm relies on the Q-network, $Q(s, a; \theta)$, the target network $Q(s, a; \theta^-)$, and the replay memory needed to sample random transitions, where $s$ is the state of the system, and $a$ are the actions already introduced. $\theta$ and $\theta^-$ are the weights of each network, which change accordingly to different updating rules. In some contexts, $\theta$ and $\theta^-$ receive the names of the current and old parameters, respectively. The replay memory stores tuples $e_t = (s_t,  a_t, r_t,  s_{t+1})$, where $s_t$ is the state of the system at time $t$, $a_t$ is the action applied at this time to drive the system, $r_t$ is the reward obtained, and $s_{t+1}$ is the state after aplying the action $a_t$. The replay memory at time $t$ comprises the previous experience tuples, $D_t=(e_0, e_1, \ldots, e_t)$. In the case of QST, the state at time $t$ is given by
\begin{equation}
s_t=(\Re{\psi_1 (t)},  \Re{\psi_2 (t)}, \ldots, \Re{\psi_N (t)}, \Im{\psi_1 (t)}, \Im{\psi_2 (t)}, \ldots, \Im{\psi_N (t)}), 
\end{equation}
where $\psi_k(t)$ is the $k$ component of the chain vector state at time $t$. 

The updating of the current parameters minimises the loss function given by
\begin{equation}
L(\theta_i) = \mathbb{E} \left[
\left(
r + \gamma \max_{a^{\prime}} Q(s^{\prime},a^{\prime}; \theta_i^-) - Q(s,a_i,\theta_i)
\right)^2
\right],
\end{equation}
where the expectation value runs over the actions $a$, the states $s$ and $s^{\prime}$ and the replay memory. We follow the notation used in Reference~\cite{Nair2015} for the weight parameters $\theta$ and $\theta^-$.  

In the jargon used in RL, an agent drives the time evolution of the state (the quantum state of the chain) by applying one of the possible actions. At time $t$, the agent selects the action accordingly to the state and the transition policy, which is a list of probabilities that the agent
takes action $a$ when it is in the state $s$, and in doing so tries to maximise the discounted reward, defined as
\begin{equation}
R_t = \sum_{i=1} \gamma^{i-1} \, r_i ,
\end{equation}
where $r_i$ is the reward at time $t_i$, and $\gamma$ is known as the discount factor. During training, the agent follows an $\epsilon$-greedy policy: with probability $\epsilon$, it selects an action uniformly at random from the action set (exploration), and with probability $1-\epsilon$, it selects the greedy action $a_t = \arg\max_a Q_\theta(s_t, a)$ (exploitation). The value of $\epsilon$ should be explicitly specified, including whether it is held constant or decayed over training episodes or time steps. During inference or deployment, performance is evaluated under a purely greedy policy (i.e., $\epsilon = 0$), unless otherwise stated, so that reported results reflect the learned policy rather than exploratory behaviour.

Appendix~\ref{ap-RL-details} contains the details about the training procedure, hyperparameters, and other definitions that shape the working of the DRL algorithm. We do not include these details since numerous works deal with this issue; see, for instance, References \cite{Sutton1998,Li2023,Zhang2018,Sivak2022,Niu2019}. 

\begin{figure}[H]
\includegraphics[width=0.9\linewidth]{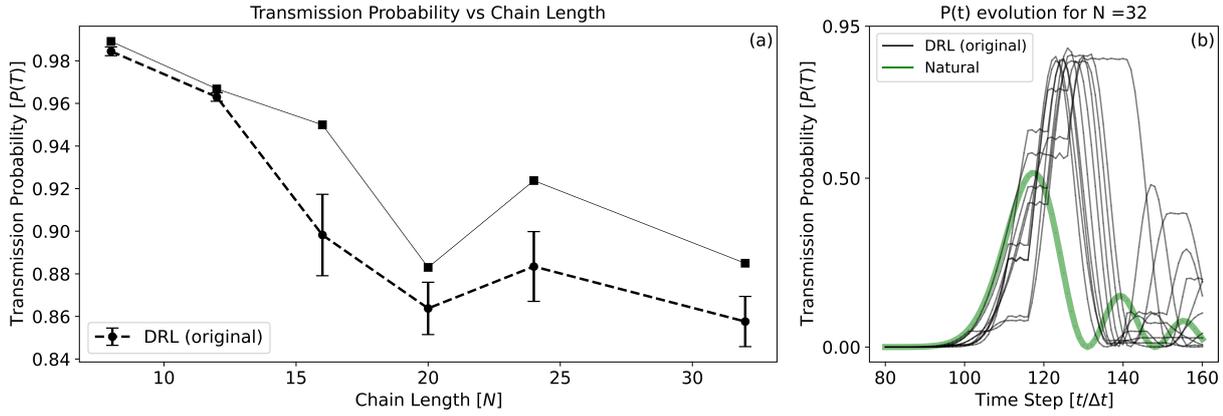}
\caption{ \label{fig-transmission-probability-drl} The figure shows (a) the maximum transmission probabilities obtained using the DQN algorithm for spin chains of different lengths, and (b) the transmission probability as a function of time for a chain of 32 qubits. The action set used to obtain the data is the one proposed by Zhang and coworkers. For the values of the hyperparameters, see the Table~\ref{tab:original-hyperparams}. While the initial state of the transmission problem is the same in all cases, that is not the case with the initial weights of the Q and target networks, even when the algorithm’s hyperparameters are kept unchanged. To test the robustness of the method, the algorithm randomly sorts the initial weights. Panel (a) presents results from 30 independent network realisations, reporting both the highest transmission probability achieved across all realisations and the average of the maximum transmission probabilities obtained in each realisation. The circular dots correspond to the averaged maximum probabilities, and the circular dots to the maximum obtained over all the realisations. The error bars correspond to the standard deviation of the data. Panel b) shows the transmission probability {\em vs } time. The black lines correspond to the forced evolution driven by the actions that achieve the maxima for the different network initial conditions, whose average and best value we plot in panel a). The green line corresponds to the autonomous, unforced time-evolution.  }
\end{figure}

Figure~\ref{fig-transmission-probability-drl}  shows the results for the transmission probability using the DRL algorithm described above. The training procedure used to obtain the control sequences that produce a good quality quantum state transfer has been described extensively in References~\cite{Zhang2018} and \cite{Fiuri2025}. The results displayed in the Figure show that the DRL algorithm is unable to find control sequences compatible with high-quality quantum state transfer, except for short to very short chains. Compare panels a) of Figures \ref{fig-transmission-probability-drl}  and  \ref{fig-fid-max-best-ga}. On the other hand, panel b) of Figure~\ref{fig-transmission-probability-drl} shows that for several realisations of the initial setting of the DRL algorithm, the time evolution of the transmission probability reaches a maximum for a time near the QSL time. Reference~\cite{Zhang2018} pointed out this phenomenon, and it is the main advantage of this method. Nevertheless, several drawbacks hinder the utility of the algorithm. First of all, the control sequence that produces the maximum value of the transmission probability may appear at any of the numerous training episodes required to reach relatively stable values for the transmission probability along the time window where the DRL algorithm trains. 

The second drawback that hinders the utility of the DRL algorithm arises from the fact that, often, after all the training episodes, the neural network is unable to provide a control sequence that effectively leads to high-quality quantum state transmission. In other words, the method finds a good control sequence at some training episode. Nevertheless, later training episodes modify the neural network parameters, effectively impairing the neural network's ability to provide good control sequences. A defective sampling of the transition probability by the neural network is usually responsible for this well-known phenomenon. Let us remember that Deep Q-networks replaced tabular methods in problems where the state and action spaces are too big, and this phenomenon appeared. Both fine-tuning the hyperparameters of the underlying network and adapting to fluctuating environments collaborate to alleviate the problem. In this case, the neural network obtained after the complete training process is robust, providing good control sequences which show resilient transmission even when the control sequence applied to the qubit chain differs from the one suggested by the trained model. Training under a fluctuating environment and the effect of poor implementation of the control sequences are the subjects of the study presented in the next Section.

\begin{figure}[H]
\begin{center}
\includegraphics[width=0.5\linewidth]{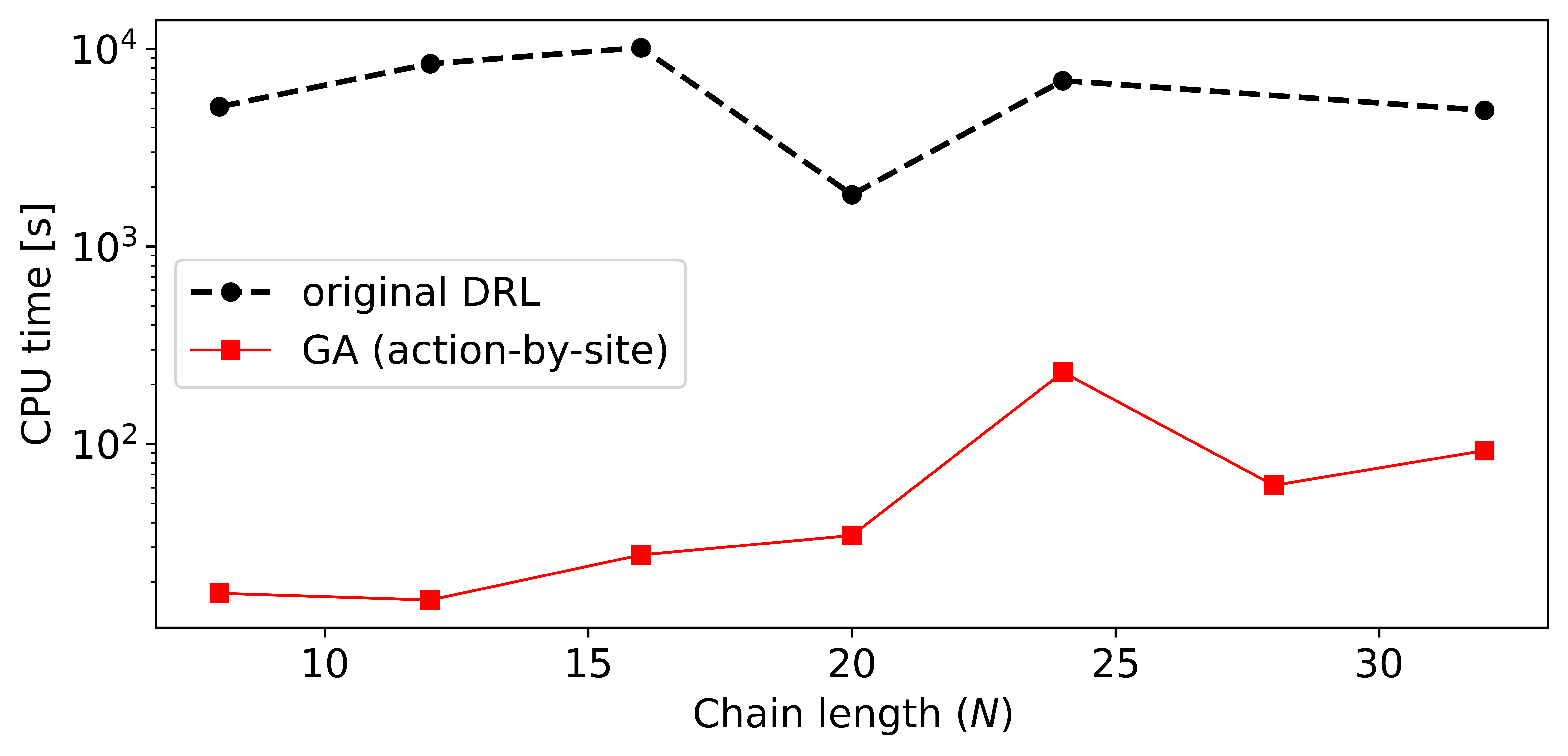}
\end{center}
\caption{The figure shows the running time required by the genetic and the DQN algorithms to reach their respective convergence criteria for qubit chains of different lengths. The black and red dots correspond to the time employed by the GA and DQN algorithms, respectively. For the GA algorithm, the criterion that halts the run is achieving the target probability of $0.99$. The criterion that halts the algorithm run in the DQN method depends on the chain length. For larger chains, the DQN algorithm halts because no improvement in the maximum value of the transmission probability appears despite further training episodes. The hyperparameters used to obtain the data in the Figure are tabulated in \ref{tab:original-hyperparams}}\label{fig-cpu-times}
\end{figure}

The third main drawback is the numerical cost of the Deep Q-network algorithm, at least when compared with the efficiency attainable by parallelising the GA. Figure~\ref{fig-cpu-times} shows the time required by the two algorithms to reach their respective convergence criteria for chains of different lengths. For the chain lengths shown (from short to moderate length), the DQN algorithm employs running times between 2 and 3 orders of magnitude larger than those required by the GA. This difference explains why, for the GA, it is possible to automate the search for suitable parameters, test different cost functions and action sets. Instead, for the DQN algorithm,  we must choose a reduced set of options to reach the best possible physical parameters and reward functions.



\section{Robustness against noisy control pulses}
\label{sec-noisy-control}

The inclusion of dynamic noise when simulating the time evolution of a forced qubit chain obeys several reasons. For instance, the dynamic noise can arise due to an imperfect implementation of the phase gates, noisy electronics, or because quantum systems are actually open. Having this in mind, and because we want to test the idea of training the DQN algorithm in a fluctuating environment, we proceed to introduce imperfect control sequences by using a random term. From the point of view of Reinforcement Learning, including a fluctuating environment is known as {\bf domain randomisation}, a technique that forces the agent to learn robust policies. The application of this technique ensures that a single instance falls within the training distribution, making the model (in our case, the DQN model) more adaptable. See, for example, References \cite{Tobin2017, Weng2019}. Closely related to the concept of robust models generated through domain randomisation is that of validation,  a process that effectively checks the robustness of the model by testing the agent through several instances. See References \cite{Gruosso2021, Wu2024}.

\begin{figure}[H]
\includegraphics[width=0.60\linewidth]{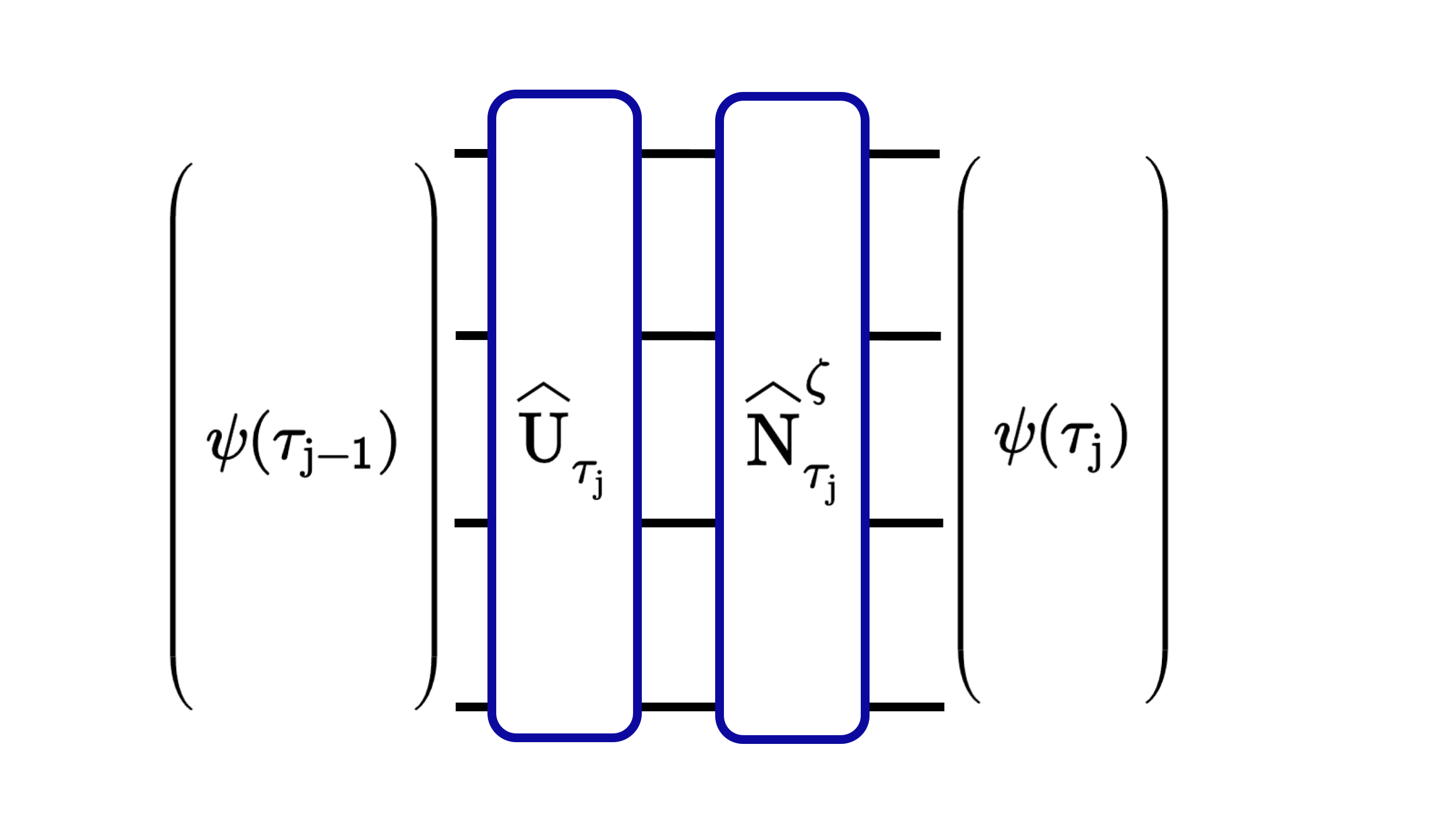}
\caption{The cartoon depicts the mechanism that introduces random noise on the quantum state of the chain. In the zero noise scenario $\psi(\tau_j) = U(\tau_j) \psi(\tau_{j-1})$., as depicted in the cartoon in Figure~\ref{fig-cartoon-circuit}, and shown in Eq.~\eqref{ec-evolution-operator}. When the system experiences noise, at each temporal interval, the evolution algorithm sorts a random variable $\zeta$. If $\zeta < p$, where $p$ is the noise occurrence probability, then the algorithm applies a random phase unitary gate to the quantum state, denoted as $N_{\tau_j}^{\zeta}$. The effect of the noise is similar to a {\em dephasing gate}.   }\label{fig-cartoon-random-phase}
\end{figure}

To add a fluctuating environment, we proceed as follows. At each temporal step and with a certain probability $p$, each component of the vector state changes as
\begin{equation}\label{ec-random-phase-evolution}
\psi_k(\tau_j) \longrightarrow \psi_i(\tau_j) \exp{(i \delta \xi_k)} ,
\end{equation}
where $\delta$ is the noise amplitude and $\lbrace \xi_k \rbrace $ are random independent variables uniformly distributed, such that $\xi_k \in \left[ -1,1\right]$. The cartoon in Figure~\ref{fig-cartoon-random-phase} depicts the time evolution of the quantum state of the chain under this updating rule. We model noise by using a noise gate $N^\zeta_{\tau_j}$ which is defined as:

\begin{equation}
N_{\tau _{j}}^{\zeta } \ =\left\lbrace
\begin{matrix}
\begin{pmatrix}
e^{i\delta\, \xi _{1}^{j}} & \dotsc  & 0\\
\vdots  & e^{i\delta \,\xi _{k}^{j}} & \vdots \\
0 & \dotsc  & e^{i\delta\, \xi _{N}^{j}}
\end{pmatrix} \  & \text{if } \zeta \ < p\\
 & \\
\mathbb{I}^{N\times N} & \ \text{if } \zeta \ \geq p
\end{matrix} \ \right. 
\end{equation}

\noindent where $\zeta$ is a random variable chosen from a uniform distribution over $\left[0,1\right]$. Note that at a given temporal step, the noise term affects all the coefficients of the vector state or none.    

Of course, there are other possible ways to introduce noise terms. In particular, the probability that a given coefficient of the vector state results modified could be chosen site by site. In this scheme, the faulty time evolution would be local, and the random term would act with a probability given in terms of the gate operation time of a qubit and a decoherence time. Since we aim to train DQN models in fluctuating environments, we will stick to the scheme given in Eq.~\eqref{ec-random-phase-evolution}. We think that the "global" scheme, where all the coefficients of the vector state change at the same temporal interval, spoils the transmission more than the site-by-site scheme, especially for values of $p$ close to unity, where the noise term acts over all the coefficients of the vector state at every interval $\tau_k$. In the global scheme, a noisy event applies random phase kicks to all amplitudes simultaneously at a given time slice, effectively producing a strong dephasing of the full state vector. In a site-by-site model only a subset of amplitudes is perturbed at each step, so phase errors partially average out and the overall degradation is typically weaker for comparable $p$ and noise amplitude. 

\begin{figure}[H]
\begin{center}
\includegraphics[width=0.45\linewidth]{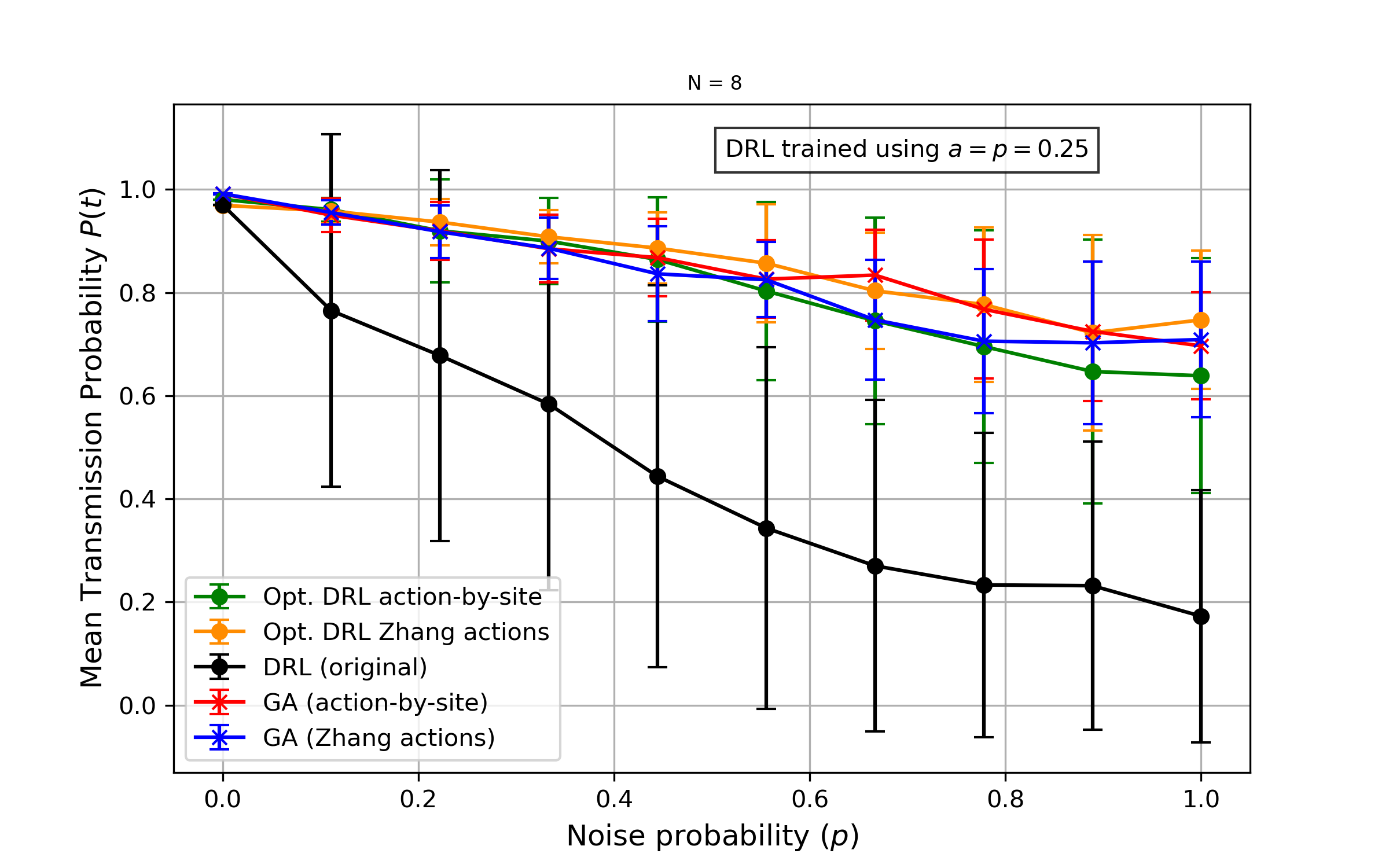}
\includegraphics[width=0.45\linewidth]{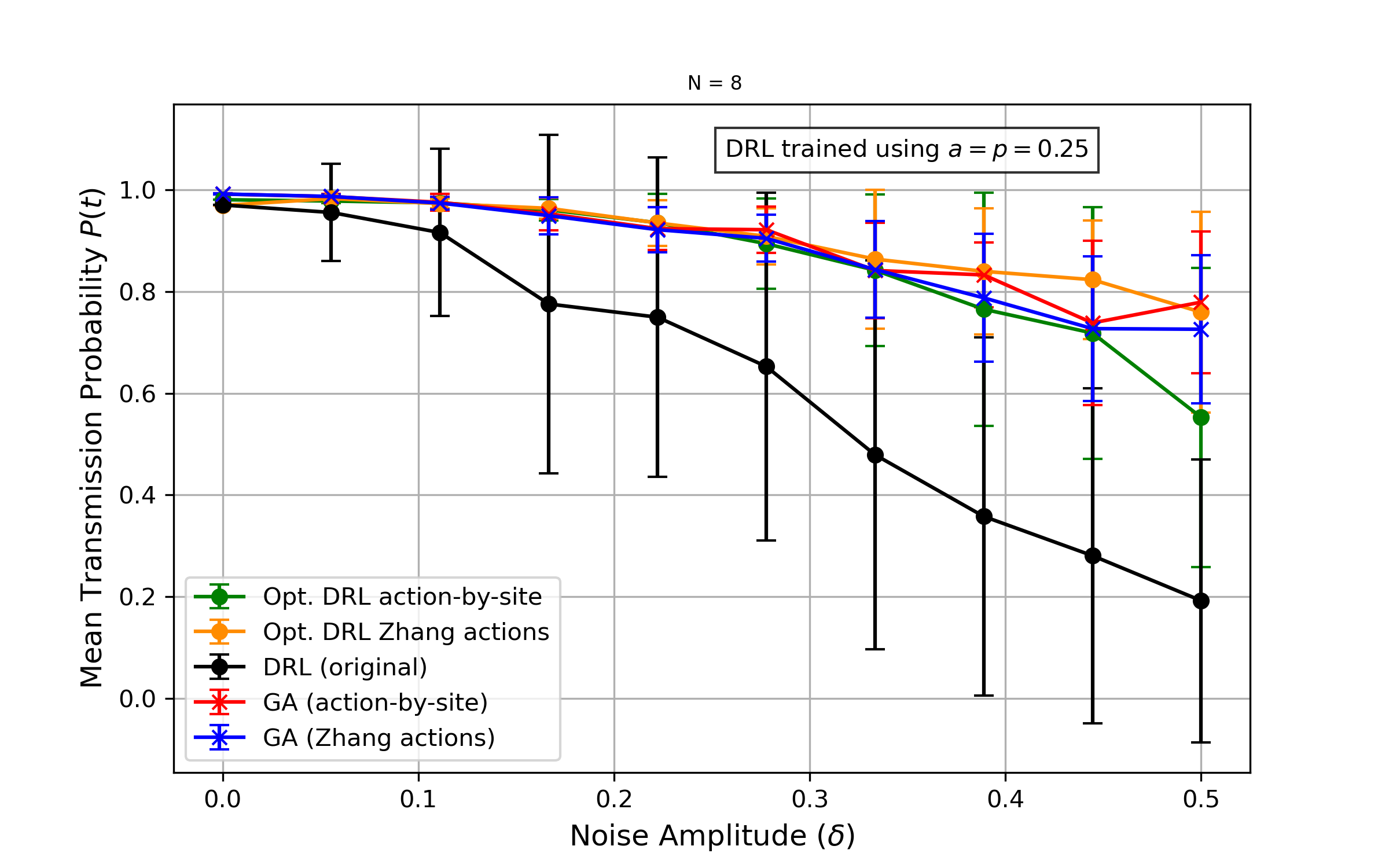}
\includegraphics[width=0.45\linewidth]{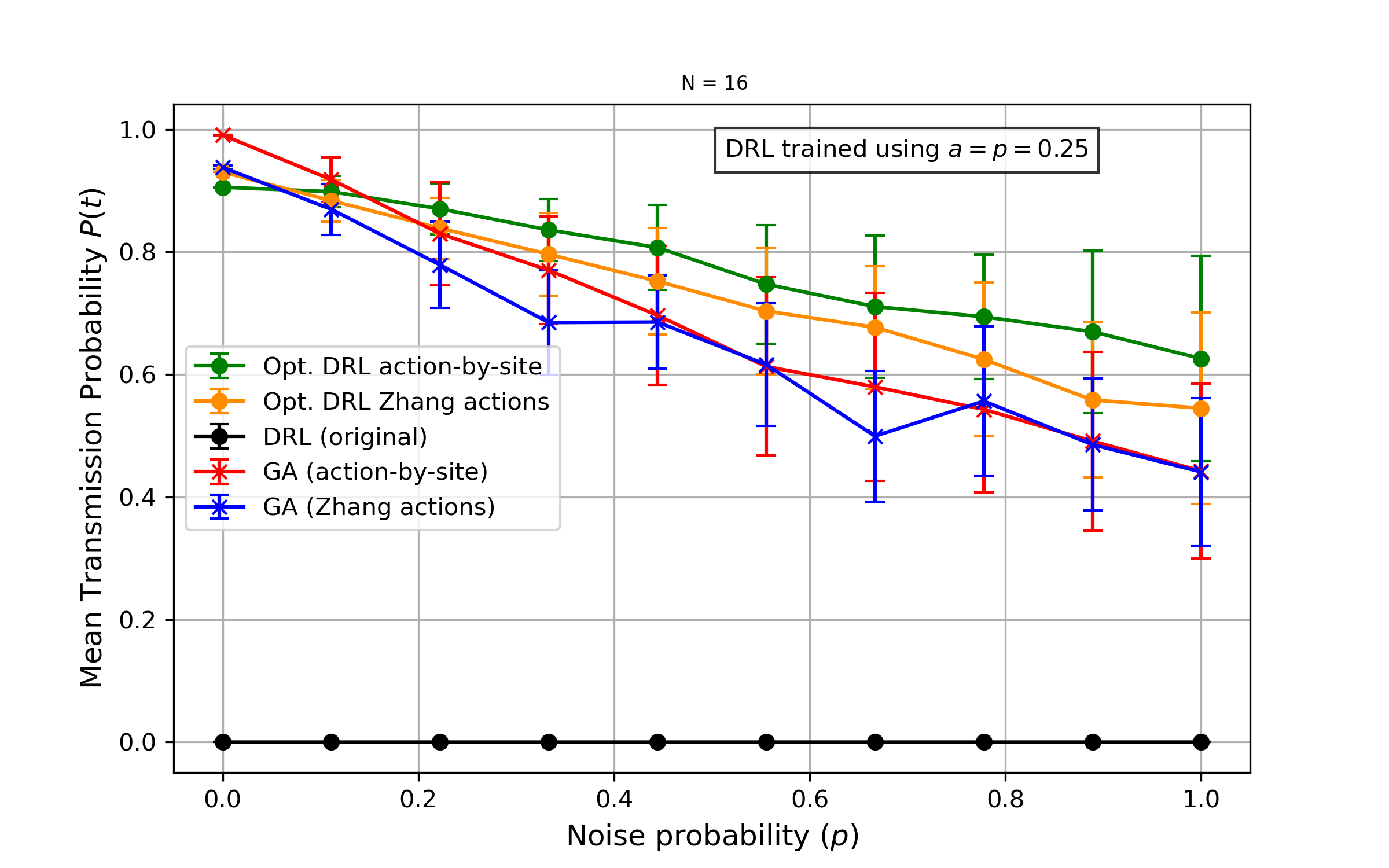}
\includegraphics[width=0.45\linewidth]{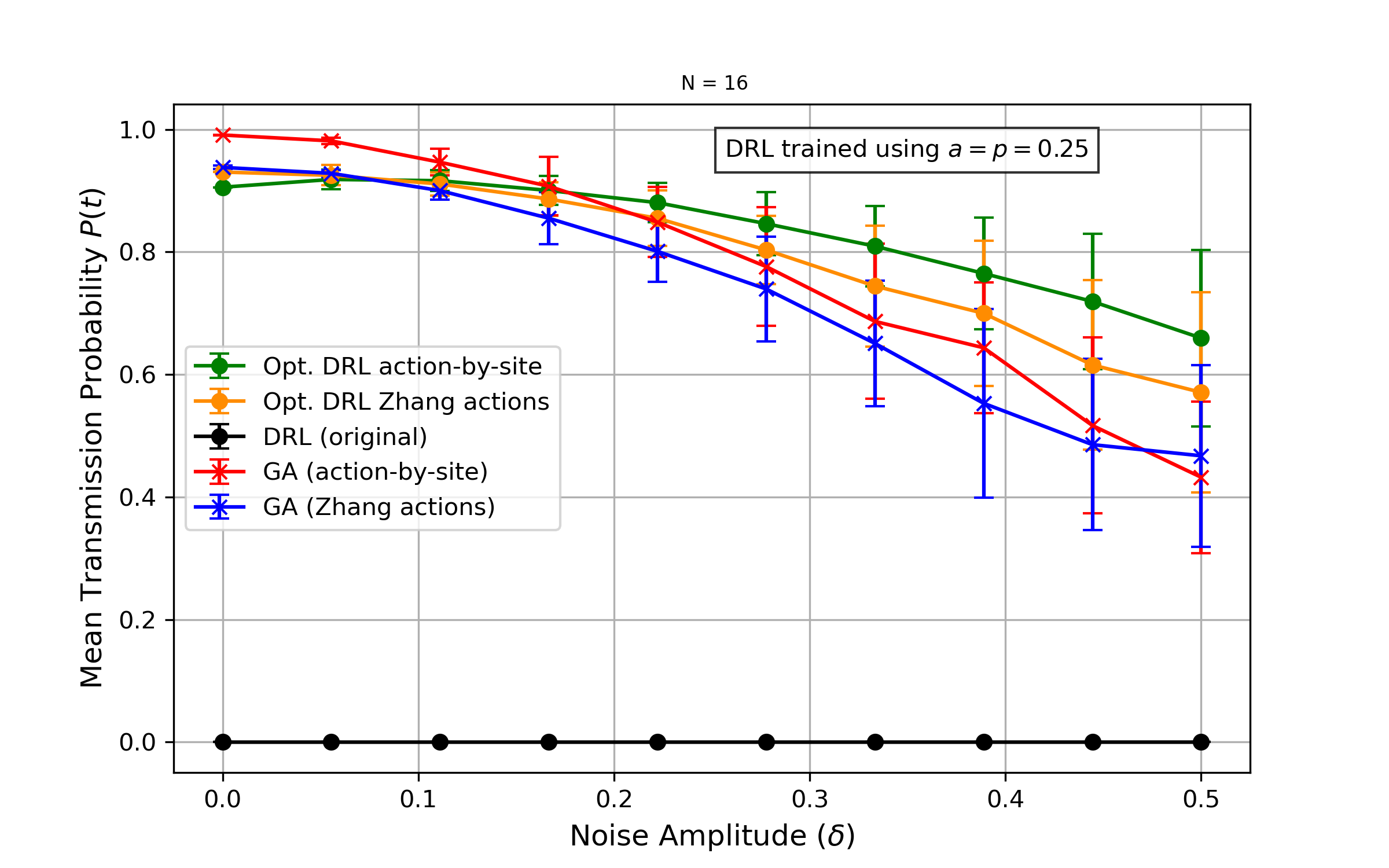}
\includegraphics[width=0.45\linewidth]{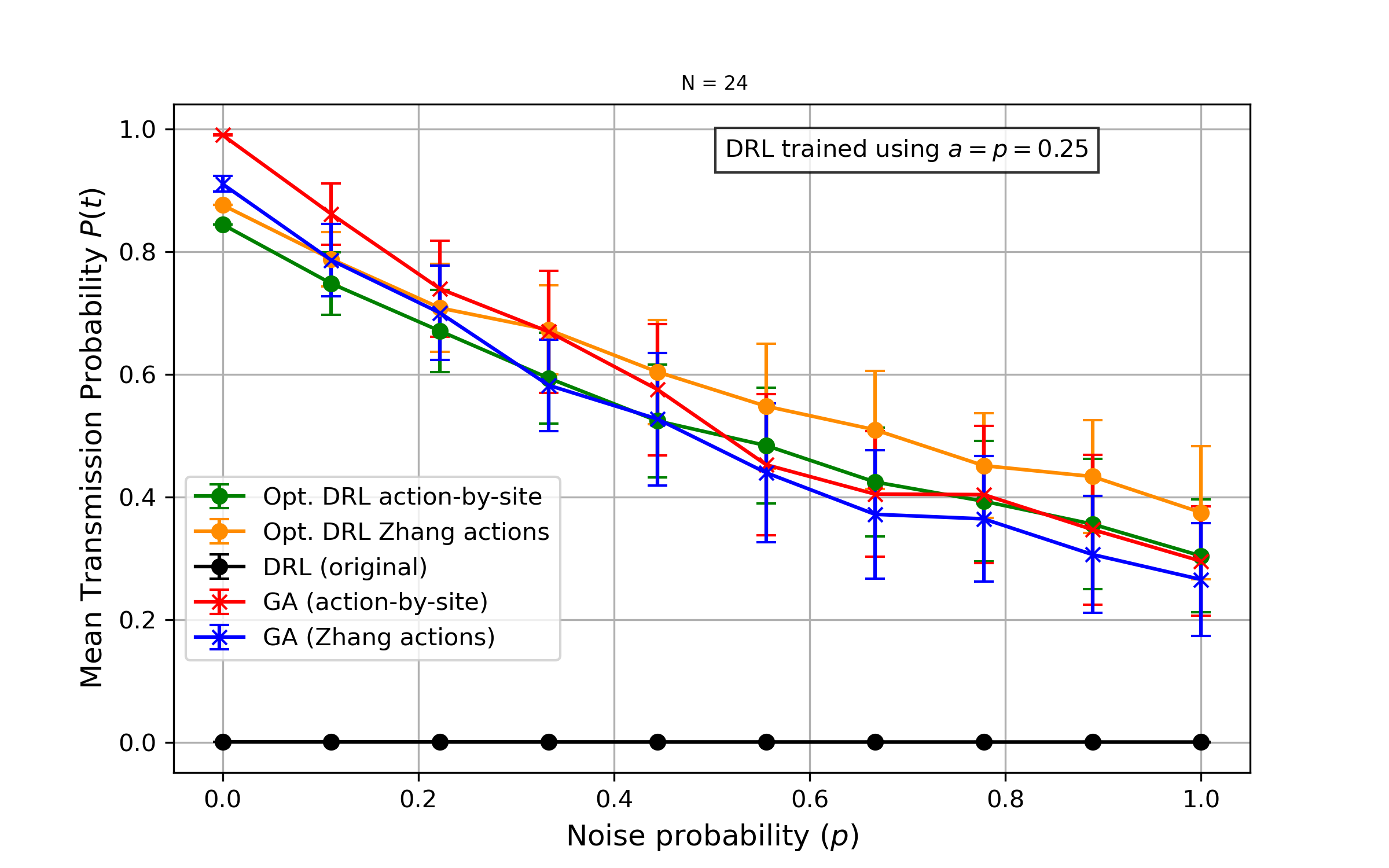}
\includegraphics[width=0.45\linewidth]{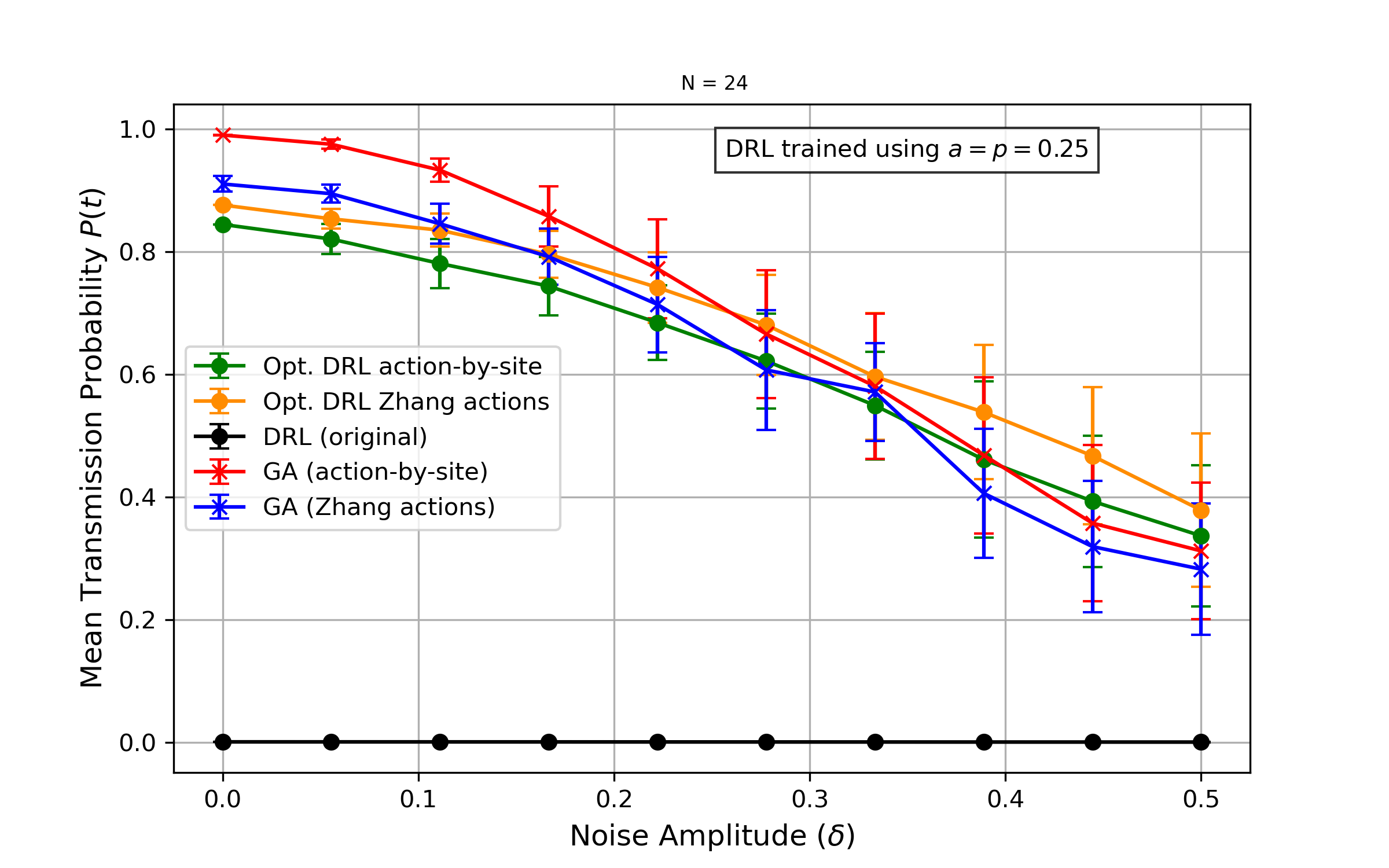}
\includegraphics[width=0.45\linewidth]{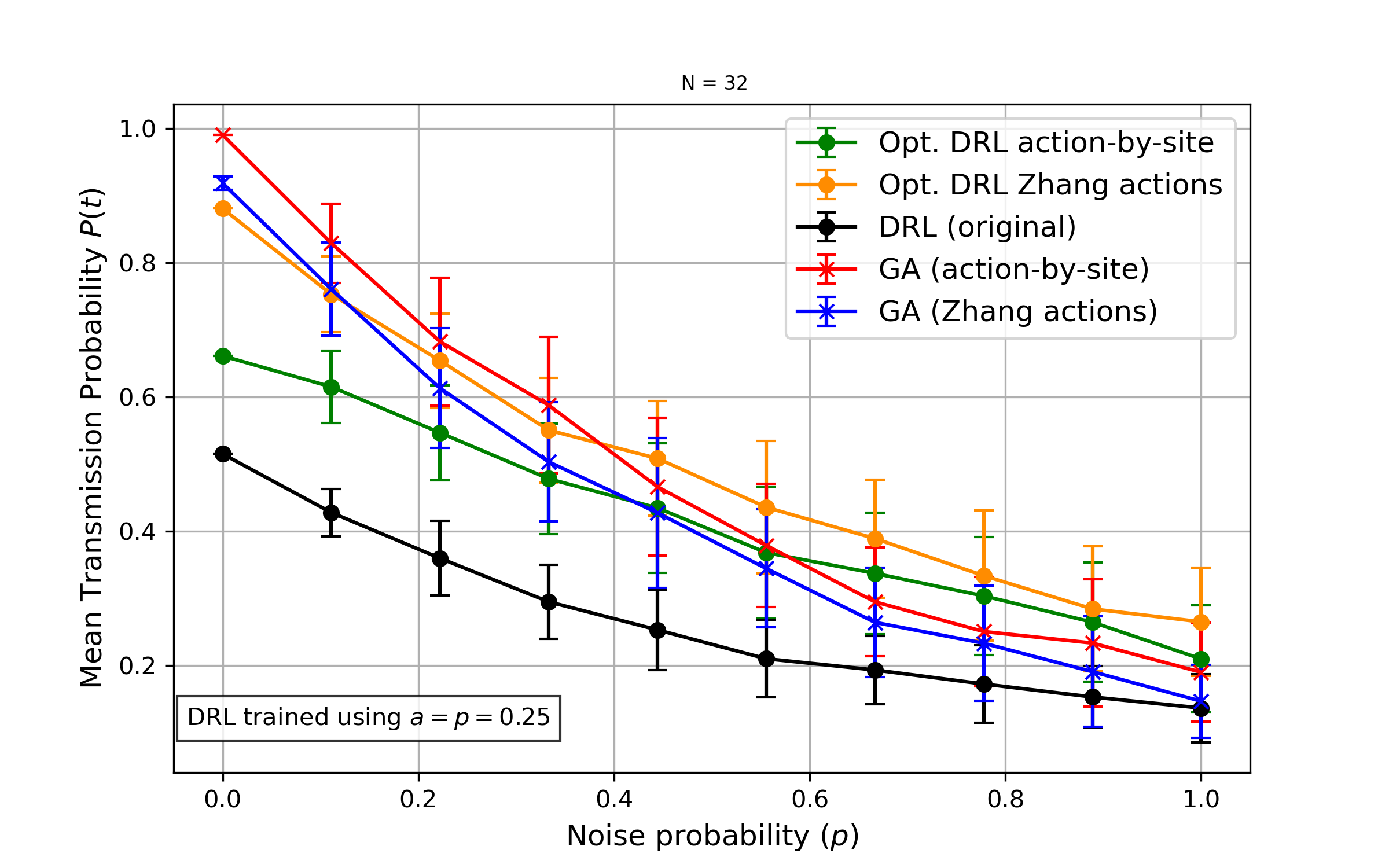}
\includegraphics[width=0.45\linewidth]{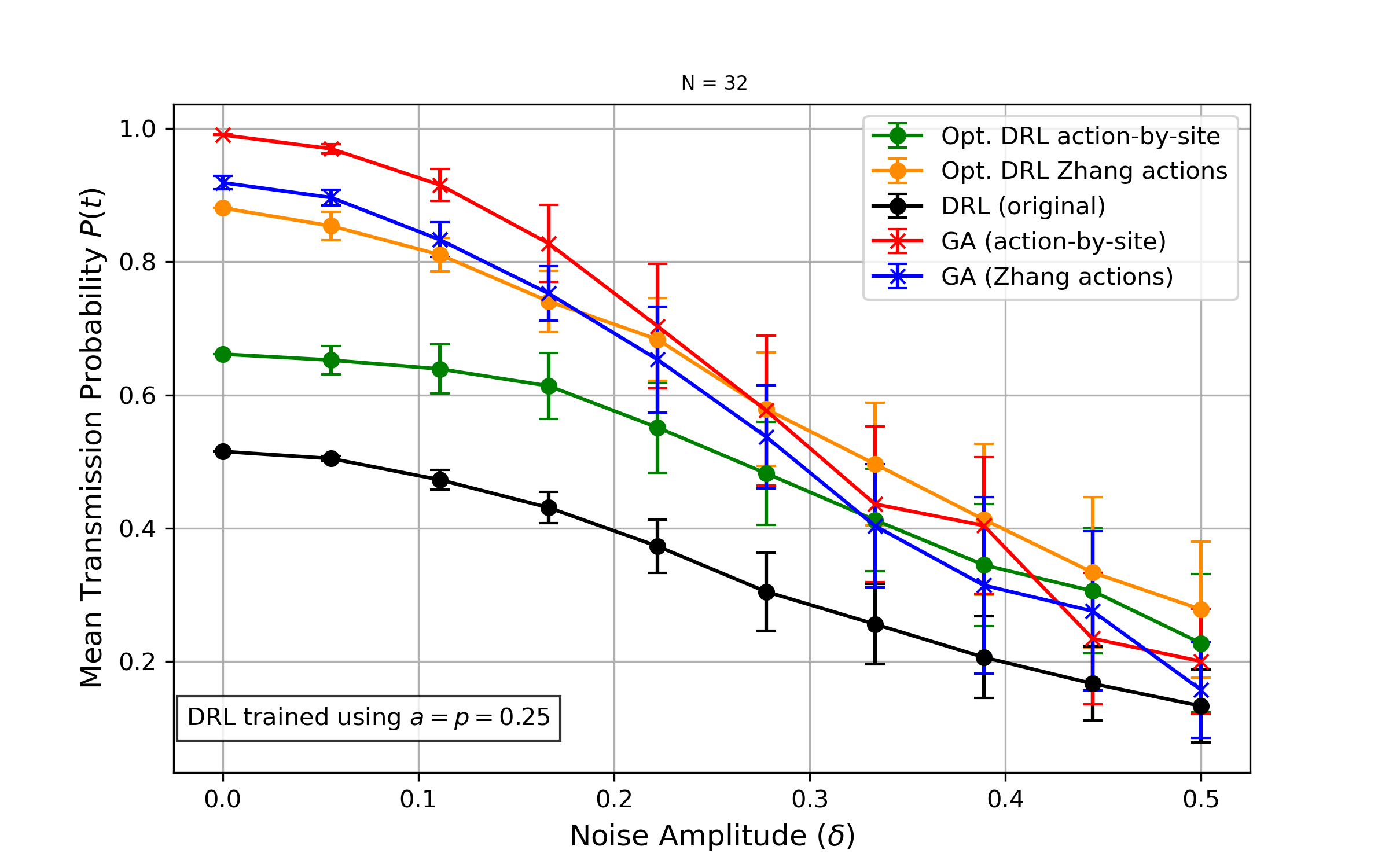}
\end{center}
\caption{The figure shows the transmission probability obtained using the validation procedure. We plot the transmission probability at a fixed amplitude in panels a), c), and e), or fixed probability in panels b), d), and f). Panels a) and b) correspond to a chain with $N=8$ qubits, c) and d) to $N=16$, and e) and f) to $N=24$, respectively. The data points correspond to the maximum transmission probability averaged over the $100$ validation runs, and the error bars to its standard deviation. The data corresponding to the DRL using actions site-by-site, DRL using the action set proposed by Zhang and coworkers, both with hyperparameters optimised, the DRL model with the hyperparameters used in Reference \cite{Zhang2018}, the GA with site-by-site actions and the action set proposed in Reference \cite{Zhang2018} correspond to the green, orange, black, red and blue square dots.  }\label{fig-figure9}
\end{figure}

To assess whether training with a fluctuating environment effectively returns a reliable model that provides control sequences producing high-quality quantum state transfer, we actualise the state $s_t$ following the prescription in Eq.~\eqref{ec-random-phase-evolution}, using fixed values of $p$ and $\delta$. After the training is complete, we validate the training by following a known procedure, which involves asking the trained agent to provide action sequences that drive the quantum system, and experiencing the effect of a noise similar to the one employed during training. 

Figure~\ref{fig-figure9} shows the transmission probability obtained using the validation procedure. We first train the DRL method in a fluctuating environment, where the noise inducing these fluctuations occurs with a probability of $ p = 0.25$ for each time interval $\Delta t$, and an amplitude of $ \delta = 0.25$. The validation procedure involves $100$ runs of the algorithm, where each run starts by initialising the quantum state in the $|\mathbf{1}\rangle$ state and asking the agent for actions to control the forced evolution. The actions drive the evolution and change accordingly to the realisation of the noise that acts during the validation runs. Note that the quantities characterising the noise acting on the validation runs, amplitude and probability of occurrence, are not the same as those used to train the DRL method. To make clearer the advantages, or drawbacks, of the training, we also include in Figure~\ref{fig-figure9} the transmission probabilities calculated using control sequences provided by the GA subjected to a dynamical noise consistent with the one used to validate the DRL method, even though the sequences obtained by the GA do not contemplate the presence of noise. 

The data shown in Figure~\ref{fig-figure9} present a multifaceted picture. First, effective transmission requires training the DQN model in the presence of noisy dynamics, together with appropriate optimisation of its hyperparameters. In the absence of such training, the validation procedure yields poor transmission probabilities, except in the limit of zero noise for short chains, as illustrated by the data shown with black dots.
Conversely, the transmission probability appears to be more robust to increases in the noise amplitude at fixed probability than to increases in the noise probability at fixed amplitude. 

Most notably, for longer chains, the control sequences generated by the GA exhibit greater resilience to dynamical noise with small amplitudes and low occurrence probabilities than those obtained using the DQN method. This advantage arises because, in the absence of noise, the GA control sequences yield higher transmission probabilities, as indicated by the data shown with green dots. However, as the noise strength increases, the averaged validation transmission probability achieved by the DQN method eventually surpasses that of the GA. This behaviour is consistent with the explicit incorporation of noisy dynamics during training and the optimisation of the hyperparameters governing the DQN method.



\section{Discussion and conclusions}
\label{sec-discussion-and-conclusions}

Up to a point, it is surprising how well the GA delivers control sequences that drive the time evolution of the chain's quantum vector state, achieving very high values of the transmission probability over an ample range of chain lengths. Although qubit chains with a hundred qubits seem a bit far-fetched right now, what our results show is that quantum systems with a Hilbert space dimension of around one or two hundred are controllable using GA generated control sequences. Moreover, the sequences are robust under weak dynamical noise.

Despite exploring several reward functions and action sets (we do not include all the combinations in the paper), the DQN method does not achieve the high-quality quantum state transfer attained by the GA. Even for short chains, the difference is not negligible and worsens when the chain length increases. A possible reason for this behaviour could lie in the fact that the GA fitness considers the effect of the control sequences along the complete time-evolution of the quantum state, from its initial condition until it arrives at the other extreme of the qubit chain, whereas the DQN method rewards local actions. Rewarding local actions is the correct strategy for numerous problems studied with DRL. Notwithstanding this, it seems insufficient to obtain control sequences from the DQN method with a performance as good as the ones produced by the GA. The studies conducted to find recognisable patterns among the actions effectively used by both methods do not suggest a practical way to bias the search in the action space toward more successful control policies. 

As the results obtained using noisy control sequences demonstrate, the performance of the control sequences generated by the DQN method will be better than that of the sequences generated by the GA for strong enough noise. Moreover, training the DQN method in a weakly noisy environment does not significantly degrade the performance of the resulting control sequences compared with those obtained under zero-noise conditions. So, the conclusion is simple. Controlling noisy quantum systems requires a method that, in the zero-noise limit, achieves control performance comparable to that of the GA. Moreover, it must be trainable under noisy conditions without substantial performance degradation to effectively control the current open, noisy quantum processors and devices.

Whereas for the GA, it is clear that the site-by-site scheme is better at controlling the time evolution of the quantum state of the qubit chain, as shown in  Figure~\ref{fig-fid-max-best-ga}, the same is not necessarily true for the DQN method when applied to noisy dynamical evolution, as shown in Panels c), d), e) and f) of Figure~\ref{fig-figure9}. An examination of the data in these panels points out that there is no single action set that is universally optimal for controlling the dynamical time evolution of the chain. Instead, the optimal choice depends on the chain length, the probability of noise occurrence and amplitude. Although the performance differences between the two action sets sometimes appear in parameter regions where the transmission probability achieved is too low to be useful, they raise questions about the utility of a given action set for different methods to obtain control sequences. 

The introduction of "softening" criteria in the fitness function is feasible, and, at least in non-controlled quantum state transfer, the transmission probabilities achievable with both types of fitness functions are comparable and very high. We are investigating a variational criterion that limits the variability and jumps commonly appearing on stepwise-like time-dependent controls. So far, we have not been able to produce an appealing proposal. 

The DQN method in particular, and DRL methods more broadly, have demonstrated remarkable success across a wide range of problems. It is therefore reasonable to expect that, in the zero-noise limit, reinforcement learning techniques can achieve very high transmission probabilities while maintaining robustness against moderate to low noise levels. Achieving improved performance with control sequences designed via DRL, however, requires at least two key ingredients. First, the algorithm must enable more effective exploration of the action-space landscape. Second, it must incorporate a multi-step policy with a reward structure that accounts for the fact that the transmission probability remains negligibly small for most of the evolution in accordance with the quantum speed limit. We are currently investigating these directions.

\subsection*{Data availability}
\noindent
The data and code supporting the findings of this study are publicly available via Zenodo at
\cite{peron_santana_source_code}.


\appendix

\section{Details about the Genetic Algorithm}
\label{ap-genetic-details}

Our Genetic Algorithm implementation utilises  PyGAD, a well-known open-source library \cite{PyGAD}. PyGAD has numerous classes that modify its behaviour accordingly to the values chosen for the hyperparameters. The implementation of the GA used in the present work employs a reduced set of hyperparameters, which we list in Table~\ref{tab:GA-hyperparams}.

\begin{table}[H]
\begin{center}
\begin{tabular}{ccccc}
\cline{1-2} \cline{4-5}
\multicolumn{2}{|c|}{\textbf{General Parameters}}                        & \multicolumn{1}{c|}{} & \multicolumn{2}{c|}{\textbf{Mutation Parameters}}                         \\ \cline{1-2} \cline{4-5} 
\multicolumn{1}{|c|}{Max. generations}      & \multicolumn{1}{c|}{1000}  & \multicolumn{1}{c|}{} & \multicolumn{1}{c|}{Mutation Type}         & \multicolumn{1}{c|}{swap}    \\ \cline{1-2} \cline{4-5} 
\multicolumn{1}{|c|}{Population size}       & \multicolumn{1}{c|}{4096}  & \multicolumn{1}{c|}{} & \multicolumn{1}{c|}{Mutation Probability}  & \multicolumn{1}{c|}{0.99}    \\ \cline{1-2} \cline{4-5} 
\multicolumn{1}{|c|}{Saturation}            & \multicolumn{1}{c|}{30}    & \multicolumn{1}{c|}{} & \multicolumn{1}{c|}{Mutated Genes}         & \multicolumn{1}{c|}{$N$}     \\ \cline{1-2} \cline{4-5} 
\multicolumn{1}{l}{}                        & \multicolumn{1}{l}{}       & \multicolumn{1}{l}{}  &                                            &                              \\ \cline{1-2} \cline{4-5} 
\multicolumn{2}{|c|}{\textbf{Parent Selection Parameters}}               & \multicolumn{1}{c|}{} & \multicolumn{2}{c|}{\textbf{Crossover Parameters}}                        \\ \cline{1-2} \cline{4-5} 
\multicolumn{1}{|c|}{Parent Selection Type} & \multicolumn{1}{c|}{'sss'} & \multicolumn{1}{c|}{} & \multicolumn{1}{c|}{Crossover Type}        & \multicolumn{1}{c|}{uniform} \\ \cline{1-2} \cline{4-5} 
\multicolumn{1}{|c|}{Parents Mating}        & \multicolumn{1}{c|}{409}   & \multicolumn{1}{c|}{} & \multicolumn{1}{c|}{Crossover Probability} & \multicolumn{1}{c|}{0.8}     \\ \cline{1-2} \cline{4-5} 
\multicolumn{1}{|c|}{Keep Elitism}          & \multicolumn{1}{c|}{409}   &                       &                                            &                              \\ \cline{1-2}
\end{tabular}

\caption{Genetic algorithm hyper-parameters. See the text for a detailed description of each one. The values shown were used to obtain the data in all the figures on the main text and Appendices. \label{tab:GA-hyperparams}}
\end{center}
\end{table}

Note that:
\begin{itemize}
\item Max generations: the maximum number of population generations. If the algorithm does not find a transmission probability larger than the target probability, and the saturation criterion is not satisfied, it halts when the number of generations reaches the Maximum value allowed.

\item Saturation: halting criterion for the algorithm. If the algorithm does not improve the maximum value of the transmission probability through the passing generations, after $30$ generations, the algorithm halts. 

\item Parent selection method: In our implementation of the GA, we use the "sss" type, which stands for {\em steady state selection}. This method chooses the best solutions (those with higher ﬁtness)  as parents, and replaces the worst solutions
with the generated offspring, ensuring that a part of the population survives and passes on to the next generation.

\item Keep elitism: This parameter sets the number of solutions with high fitness values to keep for the next generation.

\item Mutation type: the swap type interchanges two genes.

\item number of genes mutated: all the genes in the chromosome can mutate, which is equivalent to saying that a mutation can affect any pair of actions in a given control sequence. $N$ stands for the sequence length. 

\item Crossover type:  We use the uniform crossover technique, which works by
randomly selecting one of the $2$ mating parents, copying the gene from it and assigning it to the same position
in the resulting chromosome.

{\bf Set of actions proposed in  Reference \cite{Zhang2018}}. In Zhang's reference, the 16 proposed actions correspond to binary numbers. Action zero corresponds to applying no field, action one to applying a field to the first spin of the string, action two to applying a field to the second spin of the string, and action three to applying a field to the first and second spins simultaneously. The rule exploits $1=2^0$, $2=2^1$, and $3=1^0+2^1$. Following this, it is easy to see that the seventh action corresponds to applying the same field to the first, second, and third spins of the string. Actions 8 through 14 follow a similar rule, but apply to the last three spins of the string. Besides, action 15 corresponds to applying the same field to all spins.

\end{itemize}

\section{RL algorithm hyperparameters and definitions}
\label{ap-RL-details}

The DQN algorithm possesses several traits that are common in other  Reinforcement Learning algorithms. Fundamental concepts to understand how the algorithm works are the {\em training episodes}, {\em reward}, the Q and target network, etc. The cartoon in Figure \ref{fig-transmission-probability-drl} summarises some of the principal steps and concepts needed to run the algorithm. A run starts by initialising the memory and the Q-network's weights. The algorithm runs for several episodes; the quantum state evolves from the initial state (which is the same for each episode) up to a selected training time, which is larger than the transmission time. The forced evolution of the state during a single time interval results from the action chosen in that interval. The selection of the action proceeds by following what is known as an "$\epsilon$-greedy policy by the "agent". Once the state of the system (the quantum state of the chain) is updated, accordingly with the corresponding reward. All these quantities form a tuple with the experience gained $(s_t, r_t, a_t, s_{t+1})$. The algorithm stores the experience in the memory. The learning and updating of the network proceed periodically. 

The updating of the networks takes place after several time steps, where the state of the systems evolves accordingly to the actions exerted upon it. For updating, a minibatch of experiences is retrieved from the memory and used to change the weights of the network. Most of the hyperparameters of the problem determine the properties of both networks. 

Hyper-parameter tuning is an open problem in most Machine Learning problems. In order to find suitable parameters for this particular case, we used Optuna \cite{optuna}. This library allows an "educated sweep" across a selected range of values for different hyper-parameters. In particular, we optimized the values for:
\begin{itemize}
    \item $\gamma$ (reward decay), in the range: [0.95, 1]
    \item $\alpha$ (learning rate), in the range: [1E-5, 1E-2]
    \item Neurons in the first hidden layer (second hidden layer has a third of this number), in the range: [512, 4096]
\end{itemize}

\begin{figure}[H]
\includegraphics[width=0.9\linewidth]{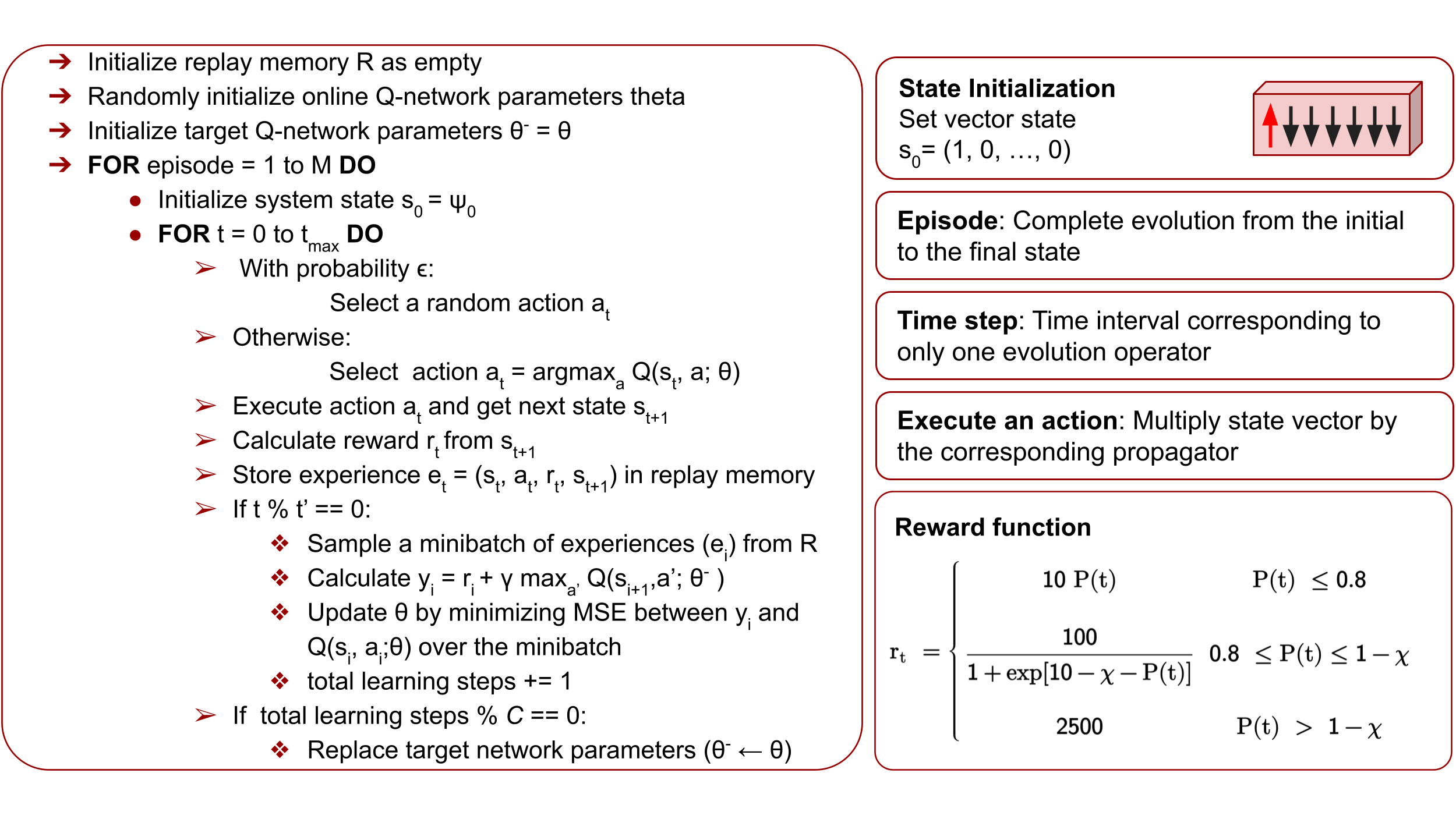}
\caption{ \label{fig-pseudo-code-drl} The cartoon depicts the main steps of the DQN algorithm, including the initialisation of the memory and the weights of the Q and target networks. The learning cycle runs over $M$ episodes,  and in each of them the algorithm forces the time evolution of the state from the initial time up to $t_{max}$, the state at the initial time is always equal to the state with a single excitation on the first site of the qubit chain. The learning process updates the weights of the Q network, whereas the weights of the target network are a copy of the weights of the Q network. The replacement of the weights of the target network occurs after the replacement period, which is much larger than the learning period. The time evolution proceeds at discrete intervals over which a given action $a_t$ acts upon the system, taking the system from $s_t$ to $s_{t+1}$ and obtaining a reward $r_t$. The reward function is a stepwise function depending on the value of the transmission probability at time $t$, $P(t)$. $\zeta$ is a threshold value; to obtain our results, $\zeta=0.05$.}
\end{figure}

We ran $32$ trials for each chain length and action set for 20000 total training episodes and 100 validation episodes. We selected the trial that provided the best average validation fidelity. The optimal hyper-parameters corresponding to the models that generated the curves on Figure \ref{fig-figure9} are shown in table \ref{tab:DRL-hyperparams}. All of them were optimized applying a 25\% noise probability and amplitude. The remaining hyper-parameters are taken from \cite{Zhang2018} and shown in Table \ref{tab:original-hyperparams}.

\begin{table}[H]
\begin{tabular}{|c|c|c|c|c|}
\hline
\textbf{Chain Length}        & \textbf{Action Set}     & \textbf{Reward Decay} ($\gamma$) & \textbf{Learning Rate} ($\alpha$) & \textbf{First Layer Neurons} \\ \hline
\multirow{2}{*}{8}  & action-by-site & 0.971                   & 0.008                    & 2417                \\ \cline{2-5} 
                    & Zhang actions  & 0.964                   & 0.001                    & 3910                \\ \hline
\multirow{2}{*}{16} & action-by-site & 0.972                   & 0.0008                   & 4062                \\ \cline{2-5} 
                    & Zhang actions  & 0.966                   & 0.0006                   & 1494                \\ \hline
\multirow{2}{*}{24} & action-by-site & 0.998                   & 0.003                    & 3219                \\ \cline{2-5} 
                    & Zhang actions  & 0.999                   & 0.0001                   & 1400                \\ \hline
\multirow{2}{*}{32} & action-by-site & 0.990                   & 0.001                    & 1502                \\ \cline{2-5} 
                    & Zhang actions  & 0.997                   & 0.0003                   & 3511                \\ \hline

Original (all lengths)           & Zhang actions  & 0.95                    & 0.01                     & 120                 \\ \hline

\end{tabular}
\label{tab:DRL-hyperparams}

\caption{Optimal hyper-parameters used to generate the curves in Figure \ref{fig-figure9}. We compare them with the ones used by the authors of \cite{Zhang2018}. Although there is not an exact way to find the best set, some general rules apply: a higher number of neurons combined with a smaller learning rate will provide more stable results. Also, better results have been found using slightly higher values of gamma, which makes sense in this setting where higher rewards are obtained in the last time steps.}

\end{table}

\begin{table}[H]
\begin{tabular}{|c|c|c|c|c|}
\cline{1-2} \cline{4-5}
\textbf{Hyper-parameter}                   & \textbf{value}       &  &\textbf{ Hyper-parameter}               & \textbf{value}                       \\ \cline{1-2} \cline{4-5} 
Minibatch size                    & 32          &  & Replay memory size            & 40000                       \\ \cline{1-2} \cline{4-5} 
Learning rate in back propagation & 0.01        &  & Neurons in 1st hidden layer    & 120                         \\ \cline{1-2} \cline{4-5} 
Reward decay ($\gamma$)           & 0.95        &  & Neurons in 2nd hidden layer & 40                         \\ \cline{1-2} \cline{4-5} 
Neurons in second hidden layer    & 40          &  & $\epsilon$-greedy rate        & 1 to 0.01 with 0.0001 decay \\ \cline{1-2} \cline{4-5} 
Learning period (t')                   & 5 timesteps &  & Number of episodes            & 50000                       \\ \cline{1-2} \cline{4-5} 
$\theta^-$  replacement period (C)                   & 200 learning episodes &  & Fidelity threshold    $ (\chi)$    & 0.0                       \\ \cline{1-2} \cline{4-5} 
\end{tabular}
\caption{Parameters used in the original DRL implementation from ref. \cite{Zhang2018}.\label{tab:original-hyperparams}}
\end{table}

\section{Action distribution for genetic algorithm solutions.}
\label{ap-histograms}

In state-of-the-art quantum processors, it is possible to apply control gates at every site of the processor. Were this not the case, there would be algorithms that would not be implementable. In other quantum systems, the local control of each qubit with every possible quantum gate is lacking. So, assuming that local phase gates are at disposal, it is worth asking if gates at all the sites are necessary to achieve very high values of the transmission probability or if there is a need to act upon the system at all at every single time step, since the unforced evolution naturally drives a single excitation located at one extreme of the chain to the other one. 

\begin{figure}[H]
    \centering
    \includegraphics[width=0.45\linewidth]{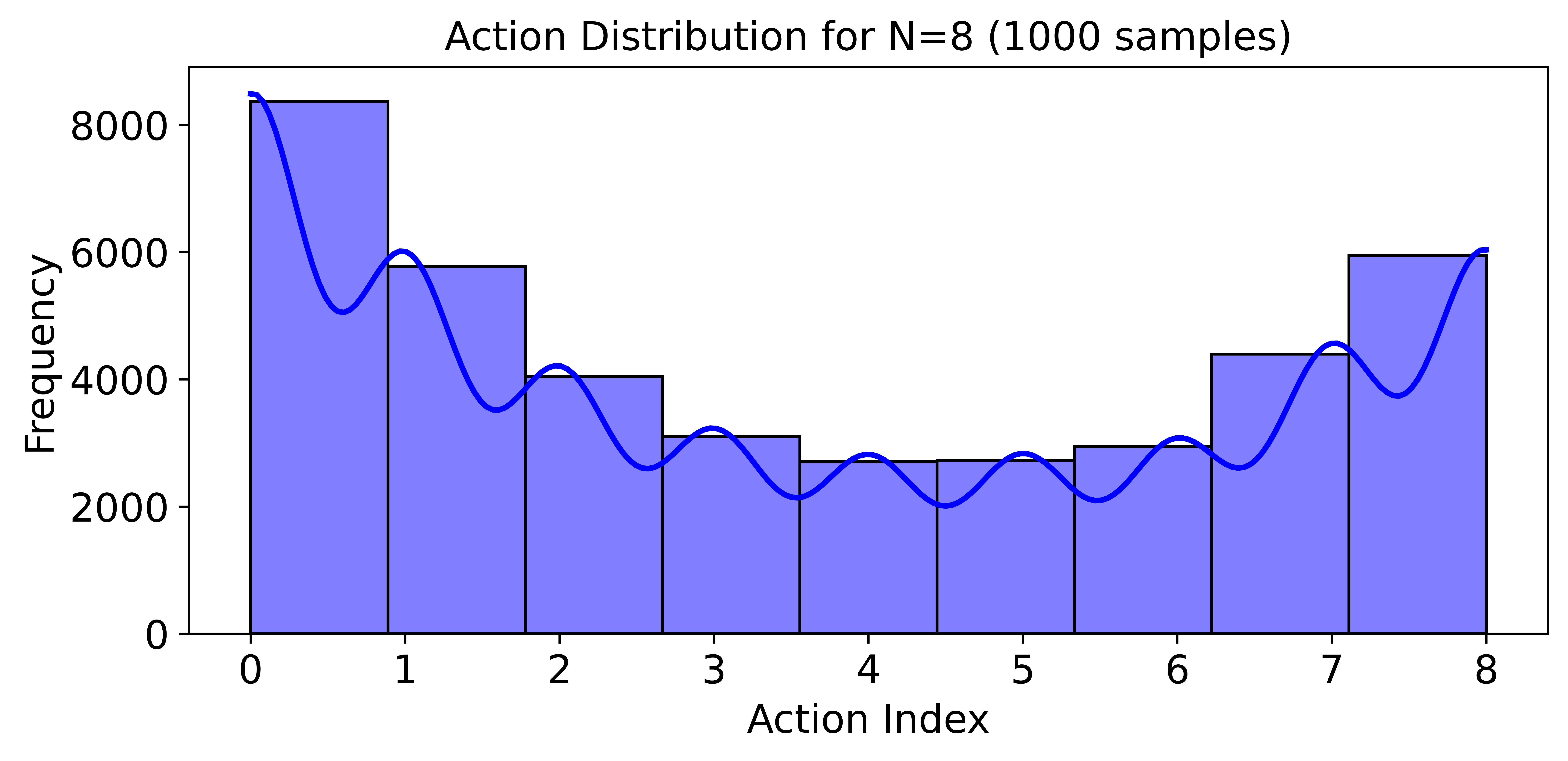}
    \includegraphics[width=0.45\linewidth]{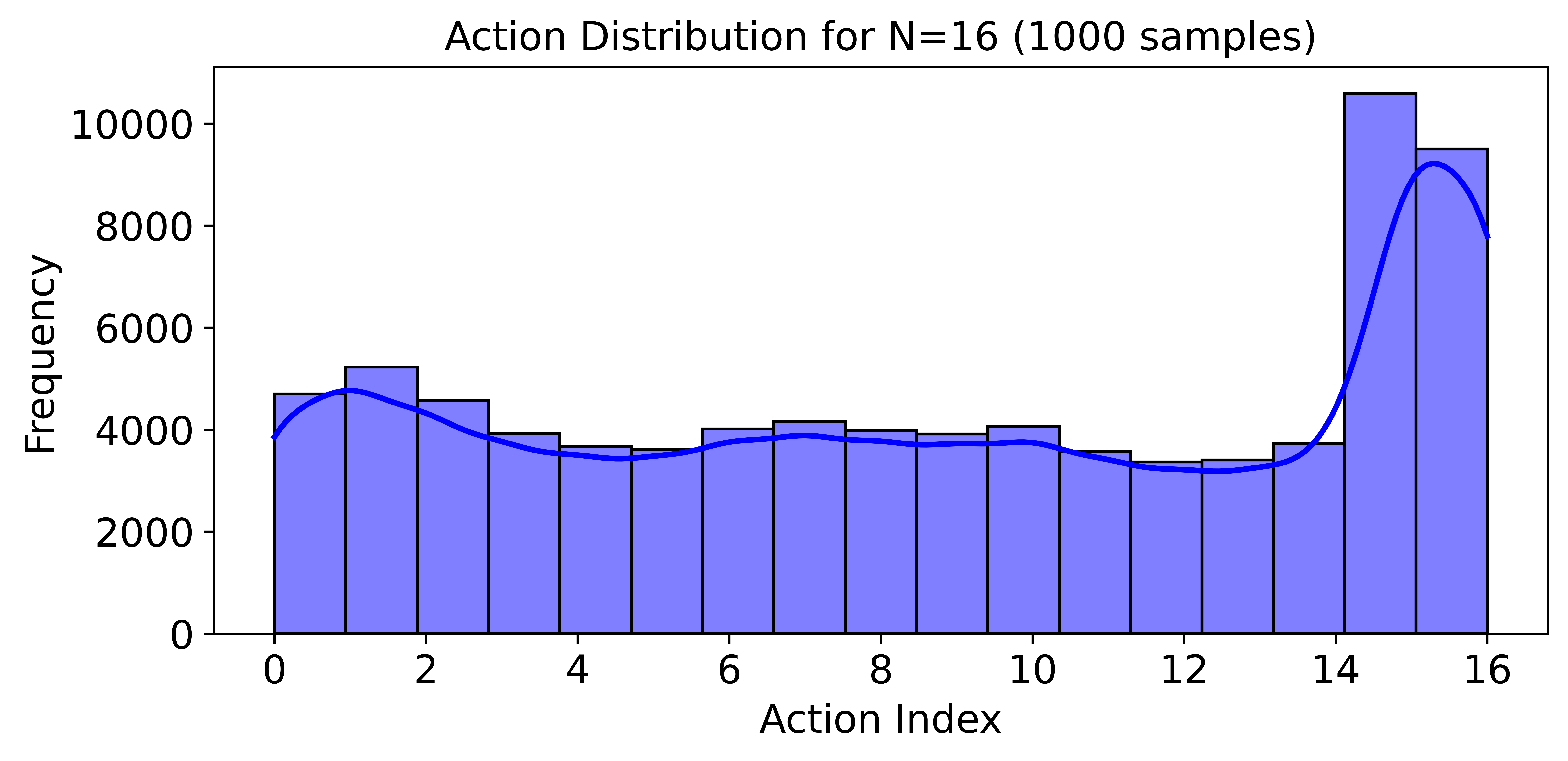}
    \includegraphics[width=0.45\linewidth]{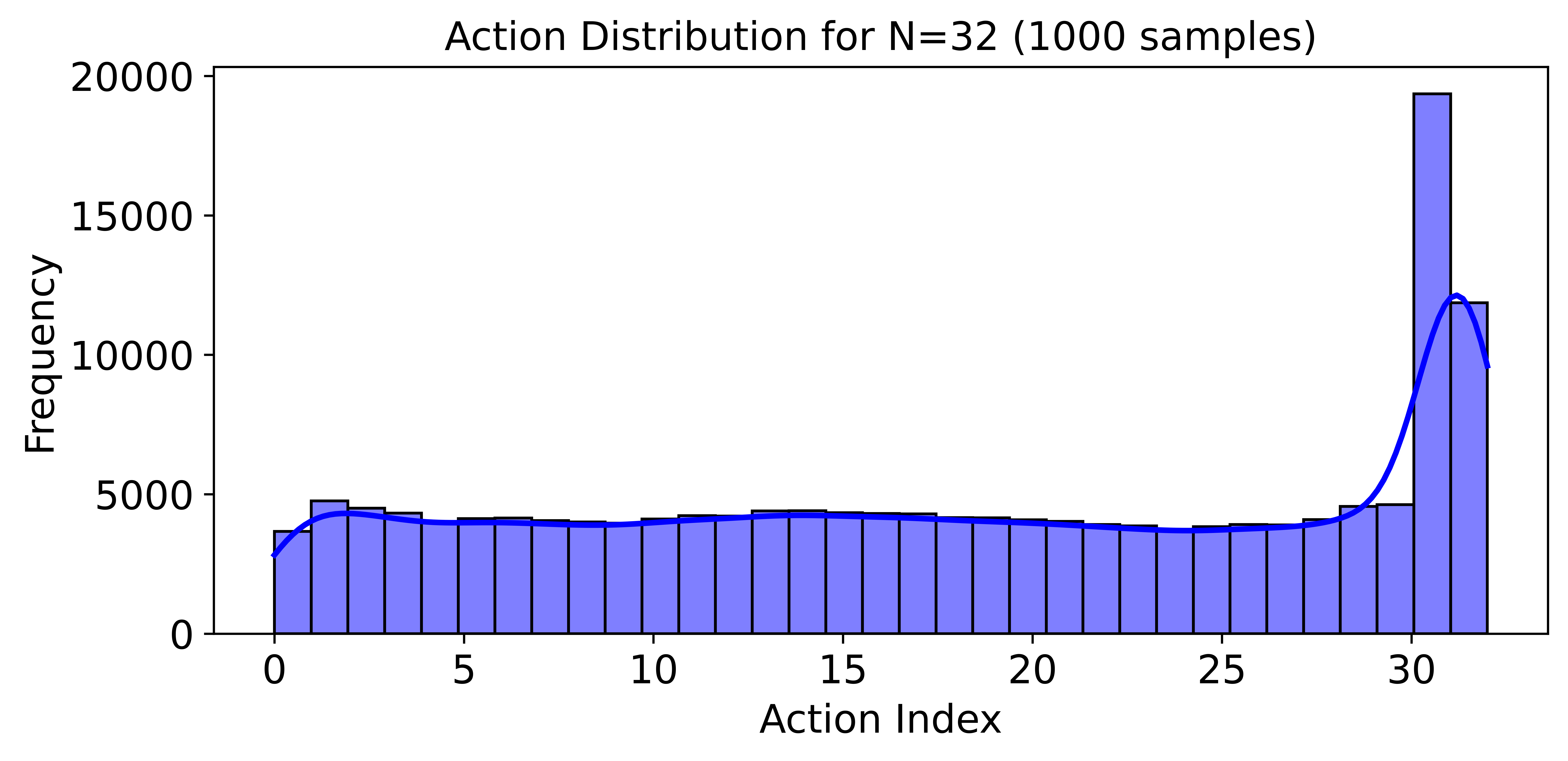}
    
    \caption{The figure shows the actions site-by-site in successful action sequences that drive the time evolution during the transfer of one excitation from one extreme of the qubit chain to the other, for different chain lengths. The data includes the actions used by $1000$ sequences, and the height of the histogram's bars corresponds to the total number of times an action acts on a given site of the chain, counting all the sequences. The "zero" action corresponds to no action exerted on the chain, not to a site. Note that the distribution is not uniform for the longer chains over their length. }
    \label{fig-histograms}
\end{figure}

To test the ideas presented in the paragraph above, we plot as histograms the action sets, site-by-site, of $1000$ successfull sequences for different chain lengths. By successful, we understand those sequences that result in transmission probabilities $P \gtrsim 0.99$. We included the parameters and hyperparameters used by the algorithm to obtain the data shown in Figure~\ref{fig-histograms} in the corresponding caption. The action denoted by zero corresponds to "no-action-at-all", whereas the other indices correspond to the site where the external field is applied. 

Note that the action zero is less and less important when the length of the chain increases, so to obtain a successful quantum state transfer, it is necessary to act upon the system with non-trivial phase gates most of the time. Moreover, acting on the qubits located in the middle of the chain is required more often than in other qubits, except for the qubits at the extremes of the chain. Interestingly, the actions at the receiving end of the chain become more relevant when the chain length increases. The action at the penultimate site of the chain occurs much more often than the other actions. We think that this is what is behind the freezing of the excitation on the last site of the chain, as observed in Figure~\ref{fig-fid-max-best-ga}. 

\section{Performance of genetic algorithm for larger qubit chains }
\label{ap-largedims}

Another interesting feature of the GA, and another advantage over DRL methods, comes from the performance of the control sequences obtained using it, since the transmission probability is not overly sensitive to increasing the system size, or equivalently, the dimension of the Hilbert space. To study this trait, we run the GA with fixed parameters and hyperparameters for chains with lengths up to $128$ sites, and for many initial populations. We use a target probability of $0.99$, which is easily attainable for chains of lengths up to $50$ qubits. For chains with length $N\leq 60$, the halting of the algorithm is a consequence of reaching the target probability, whereas for longer chains, the halting occurs by the saturation criterion. See Appendix \ref{ap-genetic-details} for the parameters used by the algorithm.

\begin{figure}[H]
    \centering
    \includegraphics[width=0.9\linewidth]{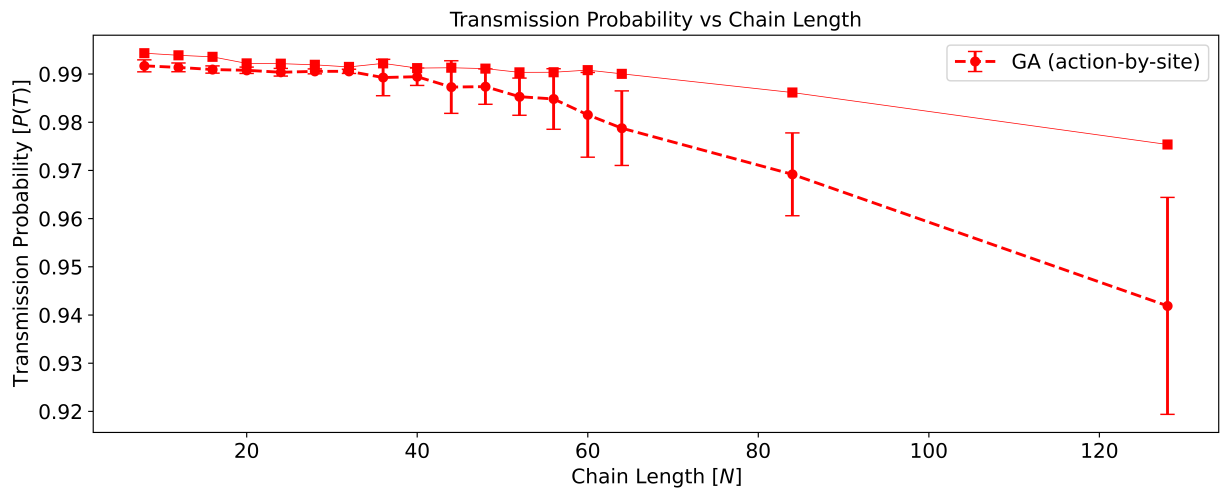}
    \caption{The Figure shows the behaviour of the transmission probability {\em vs} the chain length. The optimisation task becomes harder for increasing values of the chain length. Maintaining constant values for the number of individuals in the initial population and the other parameters means that the algorithm fails to meet the target value for the transmission probability, and the saturation criterion halts the algorithm for large enough chains. Note that for lengths $N\leq 60$, the algorithm returns transmission probabilities larger than the target value (square red dots), whereas for $N>60$, the best values do not reach this target. The averaged maximum transmission probability shows a more pronounced decay (circular dots). The increasing difficulty faced by the algorithm to attain the desired target is also signalled by the growing size of the error bar (shown using vertical segments). }
    \label{fig-larger-chains}
\end{figure}

Figure~\ref{fig-larger-chains} shows the transmission probability obtained using the GA with the hyperparameters listed in Table~\ref{tab:GA-hyperparams} and site-by-site actions. The square dots show the maximum transmission probability obtained by running the algorithm over $MM$ initial populations. On the other hand, circular dots correspond to the averaged maximum transmission probability, where the average runs over the number of initial populations. The averaged maximum transmission probability decays faster than the maximum, which is $P \sim 0.97$ for a chain with $128$ qubits. Improving this figure only requires a larger number of individuals and initial populations. Note that we obtained all the data using the same time interval, $\Delta t = 0.15$.

\bibliographystyle{unsrtnat}
\bibliography{refsga.bib}

\end{document}